\documentclass[fleqn,usenatbib]{mnras}

\usepackage[T1]{fontenc}
\usepackage{ae,aecompl}

\usepackage{graphicx}	% Including figure files
\usepackage{amsmath}	% Advanced maths commands
\usepackage{amssymb}	% Extra maths symbols
\usepackage{hyperref}
\usepackage{xcolor}     % Font colors

\usepackage{graphicx,color}	% Including figure files

\newcommand{\kms}{\textrm{km s}^{-1}}
\newcommand{\hi}{\text{H\,\sc{i}}}
\newcommand{\Msol}{\textrm{M}_{\odot}}

\newcommand{\barolo}{\textsc{3DBAROLO}~}

\newcommand{\KS}[1]{\textcolor{black}{#1}}

\title[Polar Rings]{WALLABY Pilot Survey: \textcolor{black}{the Potential Polar Ring Galaxies NGC~4632 and NGC~6156}}

\author[N. Deg et al.]{
N. Deg,$^{1}$\thanks{E-mail: nathan.deg@queensu.ca}
R. Palleske$^{2}$,
K. Spekkens$^{1,2}$,
J. Wang$^{3}$,
T. Jarrett$^{4}$,
J. English$^{5}$,
X. Lin$^{3}$,
\newauthor
J. Yeung$^{5}$,
J. R. Mould$^{6}$,
B. Catinella$^{7,8}$, 
H. D\'{e}nes$^{9}$,
A. Elagali$^{10,11}$,
B.~-Q. For$^{7,8}$,
\newauthor
P. Kamphuis$^{12}$,
B.S. Koribalski$^{13,14}$,
K. Lee-Waddell$^{7,15,16}$,
C. Murugeshan$^{7,8}$,
S. Oh$^{17}$,
\newauthor
J. Rhee$^{7,8}$,
P. Serra$^{18}$,
T. Westmeier$^{7}$,
O.I. Wong$^{15,7,8}$,
K. Bekki$^{7}$,
A. Bosma$^{19}$,
\newauthor
C. Carignan$^{4,20,21}$,
B.W. Holwerda$^{22}$,
N. Yu$^{23}$
\\
$^{1}$Department of Physics, Engineering Physics, and Astronomy, Queen's University, Kingston, ON, K7L 3N6, Canada\\
$^{2}$Department of Physics and Space Science, Royal Military College of Canada, P.O.\ Box 17000, Station Forces Kingston ON K7K~7B4, Canada\\
$^{3}$Kavli Institute for Astronomy and Astrophysics, Peking University, Beijing 100871, China \\
$^{4}$Department of Astronomy, University of Cape Town, Private Bag X3, Rondebosch 7701, South Africa\\
$^5$Department of Physics and Astronomy, University of Manitoba, Winnipeg, Manitoba, R3T 2N2, Canada\\
$^{6}$Centre for Astrophysics \& Supercomputing, Swinburne University, P.O.Box 218, Hawthorn, Vic 3122, Australia\\
$^{7}$ ICRAR - The University of Western Australia, 35 Stirling Highway, Crawley WA 6009, Australia \\
$^{8}$ ARC Centre of Excellence for All Sky Astrophysics in 3 Dimensions (ASTRO 3D), Australia\\
$^{9}$ School of Physical Sciences and Nanotechnology, Yachay Tech University, Hacienda San Jos\'{e} S/N, 100119, Urcuqu\'{i}, Ecuador  \\
$^{10}$ Minderoo Foundation, 171 - 173 Mounts Bay Road, Perth, WA 6000, Australia\\
$^{11}$ School of Biological Sciences, The University of Western Australia, 35 Stirling Highway, Crawley, WA 6009, Australia \\
$^{12}$ Ruhr University Bochum, Faculty of Physics and Astronomy, Astronomical Institute (AIRUB), 44780 Bochum, Germany \\
$^{13}$ CSIRO Space Science and Astronomy, Australia Telescope National Facility, P.O. Box 76, NSW 1710, Australia \\
$^{14}$ School of Science, Western Sydney University, Locked Bag 1797, Penrith, NSW 2751, Australia \\
$^{15}$ CSIRO Space and Astronomy, PO Box 1130, Bentley WA 6102, Australia\\
$^{16}$ ICRAR - Curtin University, Bentley, WA 6102, Australia\\
$^{17}$ Department of Physics and Astronomy, Sejong University, Seoul 05006, Korea\\
$^{18}$ INAF - Osservatorio Astronomico di Cagliari, Via della Scienza 5, I-09047 Selargius (CA), Italy\\
$^{19}$Aix Marseille Univ, CNRS, CNES, LAM, Marseille\\
$^{20}$D\'{e}partement de physique, Universit\'{e} de Montr\'{e}al, Complexe des sciences MIL, 1375 Avenue Th\'{e}r\`{e}se-Lavoie-Roux Montr\'{e}al, \\ Qc, Canada H2V 0B3 \\
$^{21}$Laboratoire de Physique et de Chimie de l’Environnement, Observatoire d’Astrophysique de l’Université \\ Ouaga I Pr Joseph Ki-Zerbo (ODAUO), BP 7021, Ouaga 03, Burkina Faso\\
$^{22}$University of Louisville, Department of Physics and Astronomy, 102 Natural Science Building, 40292 KY Louisville, USA\\
$^{23}$National Astronomical Observatories, Chinese Academy of Sciences, 20A Datun Road, Chaoyang District, Beijing 100871, People's Republic of China
}
\date{Accepted XXX. Received YYY; in original form ZZZ}

\pubyear{2022}

\begin{document}
\label{firstpage}
\pagerange{\pageref{firstpage}--\pageref{lastpage}}
\maketitle

% Abstract of the paper
\begin{abstract}
We report on the discovery of two \textcolor{black}{potential} polar ring galaxies \textcolor{black}{(PRGs)} in the WALLABY Pilot Data \textcolor{black}{Release} 1 (PDR1).  These \KS{untargetted detections}, cross-matched to NGC 4632 and NGC 6156, are some of the first \textcolor{black}{galaxies where the \hi\ observations show two distinct components.}
%first polar ring galaxies where the apparent polar structures have been found in the \hi\ observations rather than through their stellar structures.  
We used the \textsc{iDaVIE} virtual reality software to separate the anomalous gas from the galactic gas and find that the anomalous gas comprises $\sim50\%$ of the total \hi\ content of both systems.  We have generated plausible \KS{3D} kinematic models for each galaxy assuming that the rings are circular and inclined at $90^{\circ}$ to the galaxy bodies. \textcolor{black}{These models show that the data are consistent with PRGs, but do not definitively prove that the galaxies are PRGs.}  \KS{By projecting these models at different combinations of main disk inclinations, ring orientations, and angular resolutions in mock datacubes, we} have further investigated the detectability of similar \textcolor{black}{PRGs} in WALLABY.  \textcolor{black}{Assuming that these galaxies are indeed PRGs, the} detectability fraction, combined with the size distribution of WALLABY PDR1 galaxies, implies an incidence rate of $\sim 1\%-3\%$.
%, which is much higher than the canonical rate of $0.1\%$.
If this rate holds true, the WALLABY survey will \textcolor{black}{detect} hundreds of new polar ring galaxies.
\end{abstract}

% Select between one and six entries from the list of approved keywords.
% Don't make up new ones.

%%%%%%%%%%%%%%%%%%%%%%%%%%%%%%%%%%%%%%%%%%%%%%%%%%

%%%%%%%%%%%%%%%%% BODY OF PAPER %%%%%%%%%%%%%%%%%%

\section{Introduction}

\textcolor{black}{Polar ring galaxies (PRGs) -- systems which exhibit a ring or disk of material oriented perpendicular to the main disk -- are some of the most visually striking objects in the Universe. An extreme among the variety of kinematically-misaligned structures detected in galaxies that includes warps and counter-rotating disks \citep{Serra2014}, PRGs hold important clues for galaxy structure and evolution that range from constraining how galaxies interact \citep[e.g.\ ][]{Bekki1997, Bournaud2003,Reshetnikov1997} and accrete their gas \citep[e.g.\ ][]{Maccio2006,Brook2008, Khoperskov2021}, to probing the shapes and distributions of the dark matter halos in which they reside \citep[e.g.\ ][]{Sackett1990,Combes1996,Khoperskov2014}.}

%Indeed, PRGs could b, and other peculiar galaxies  where significant components of the galaxy have a different angular momentum than the host galaxy.

%These are galaxies where there is material that is inclined by $\sim90^{\circ}$ to the main galaxy body.  
%For PRGs, the material is in a ring, but there are also polar disk galaxies, where the material is a full extended disk \citep{Brook2008,Dzudzar2021}.  The origin of these structures provide clues to galaxy formation and star formation \citep{Egorov2019}.  Moreover, due to their orientation, they break the classic disk-halo degeneracy and allow for detailed studies of the dark matter content of their host \citep{Combes1996,Khoperskov2014}.  }

Since the discovery of the first putative stellar polar structures around nearby galaxies \citep{Sandage1961,Schecter1978}, there have been a number of attempts to measure their incidence in large optical surveys. \citet{Whitmore1990} searched for polar structures around S0 galaxies and generated a catalogue of 157 PRG candidates. \textcolor{black}{They confirmed that PRGs require kinematic followups to determine that the polar material is indeed rotating about the inner galaxy with the same center and with a \textit{large} inclination with respect to galaxy's plane.  PRG candidates are objects where these sorts of kinematic followups have not yet been completed.}  Based on the fraction of galaxies searched that exhibit potential polar rings (0.5\%) and the geometric detectability of these structures depending on their sky projection, \citet{Whitmore1990} estimated 5\% of all S0 galaxies may be PRGs.  \citet{Moiseev2011} used Galaxy Zoo  \citep{Lintott2008} classifications of Sloan Digital Sky Survey (SDSS) images to find 275 PRG candidates, while \citet{Reshetnikov2019} also mine SDSS to identify 31 new PRG candidates. These searches suggest that, in contrast to milder kinematic misalignments \citep[e.g.\ ][]{GarciaRuiz2002,Ann2006,Ristea2022}, stellar PRGs are rare within the galaxy population as a \KS{whole: their incidence is on the order of $\sim10^{-3}$ \citep{Reshetnikov2011}, and polar structures around red systems are about twice as common as those around blue systems \citep{Smirnov2022}.} 
%\textcolor{black}{A recent Despite their rarity, the sample size of PRGs is now large enough to attempt at measuring their luminosity function \citep{Smirnov2022}.}

It is worth noting that relatively few PRG candidates have been modelled in detail to confirm that they are likely to host polar structures.
\textcolor{black}{Moreover such modelling may indicate that the stellar structure is more consistent with being an extreme warp than a truly distinct ring \citep[e.g.][]{Jorza2009}.  Such a distinction is largely a matter of semantics given that both extreme symmetric warps, polar rings/disks, and inclined rings/disks are part of the larger continuum of symmetric kinematically misaligned objects.  These likely have similar origins and may form an evolutionary sequence (i.e. PRGs and inclined rings may transform into warps;  \citealt{Brook2008}).}
%\citep[e.g.][]{Jorza2009}; this implies that their true incidence may be much lower than the rates above suggest.  \textcolor{black}{}

\textcolor{black}{Atomic hydrogen (\hi) structures exhibiting extreme kinematic misalignments relative to their host galaxies have also been discovered.  \KS{Many early studies targetted stellar PRGs to look for \hi\ counterparts, finding co-located gas rings} \citep[e.g.\ ][]{vanGorkom87} or disk \citep{Brook2008,Dzudzar2021}.  %In some cases the \hi\ ring/disk is the dominant gaseous component. 
Similarly, \citet{DeRijcke2013} found a possible \hi\ ring about the FCC046 dwarf elliptical galaxy in the Fornax cluster, while \citet{Bait2020} found an offset \hi\ ring around the massive quiescent galaxy AGC~203001.  Another interesting example is the irregular galaxy NGC 6822, where 
 the gas is coincident with a young stellar disk that is inclined at $~\sim60^{\circ}$ to an extended spheroid with an older population \citep{Demers2006,Zhang2021}.  }
 
 \KS{While gas may be found around stellar PRGs during followup observations, there are also systems in which polar \hi\ components are seen without a corresponding stellar counterparts down to SDSS depths. For example, \citet{Stanonik2009} found a single \hi\ disk perpendicular to the stellar disk of the void galaxy SDSS J102819.24+623502.6., while the \hi\ ring of the SB0 galaxy NGC 4262 coincides with a series of UV-bright knots \citep{Buson2011}.}  \textcolor{black}{While these are individual galaxies, larger \hi\ surveys are \KS{also uncovering more objects} with single \hi\ components that are misaligned from the stellar structures.  \citet{Serra2012} examined the \hi\ content of 166 early-type galaxies from the ATLAS$^{\rm{3D}}$ survey \citep{Cappellari2011}.  They found three galaxies with single-component \hi\ disks that have position angles at $\sim 90^{\circ}$ to the stellar components. \KS{While no stellar counterpart is seen with the \hi\, it is worth} noting that the optical data may not yet be deep enough to see such low surface brightness features.}

 %\textcolor{black}{While gas may be found around stellar PRGs during followup observations, there are a number of galaxies where polar \hi\ is seen without a corresponding stellar feature.  \citet{Stanonik2009} found a single \hi\ disk with a position angle at $\sim90^{\circ}$ to the stellar disk of SDSS J102819.24+623502.6.  This galaxy is particularly interesting as it is found in a void.  Another galaxy found to have an \hi\ ring is the SB0 galaxy NGC 4262 \citep{Buson2011}.  This ring coincides with an outer ring of UV bright knots.  While these are individual galaxies, larger \hi\ surveys are seeing more objects with single \hi\ components that are misaligned from the stellar structures.  \citet{Serra2012} examined the \hi\ content of 166 early-type galaxies from the ATLAS$^{\rm{3D}}$ survey \citep{Cappellari2011}.  They found three galaxies with single component \hi\ disks that have position angles at $\sim 90^{\circ}$ to the stellar components.  Beyond the specific \hi\ PRGs, they found that roughly half of all the \hi\ content show a degree of misalignment from the stellar structures.  In all these cases, no stellar feature is seen with the \hi\, but it is worth noting that the optical data may not yet be deep enough to see such low surface brightness features.}

%In other cases, anomalous \hi\ is found without obvious stellar features.  For instance, observations of the irregular galaxy NGC 6822 have revealed an \hi\ disk coincident with a \citet{Demers2006} found }

\textcolor{black}{In addition to these single component \hi\ polar structures, there are a number of galaxies where the \hi\ shows multiple components.  \citet{Schiminovich2013} observed the elliptical `shell' galaxies Arp 230 and MCG -5-7-1, \KS{identifying and kinematically modelling both inner and outer \hi\ components.}  Translating their inclination and position angle fits for the inner and outer disks suggests that the Arp 230 outer structure is inclined at $\sim 60^{\circ}$ to the inner disk. For MCG -5-7-1, the inner and outer structures are less distinct kinematically, which may be due to the observations of that galaxy having a lower signal-to-noise ratio ($S/N$).  More recently, \citet{Cao2022} found two galaxies in the Mapping Nearby Galaxies at Apache Point Observatory (MaNGA; \citealt{Bundy2015}) survey \KS{harbouring inner and outer gaseous disks, that are both misaligned from the stellar component, in} a sample of 84 galaxies with large dispersions in their gaseous kinematic position angles but with well-measured stellar position angles. In this particular case they use H$\alpha$ observations of the ionized gas rather than \hi.}

In this paper, we report on two \textcolor{black}{galaxies from the WALLABY Pilot Data Release 1 \citep[PDR1,][]{SoFiARelease}, NGC 4632 and NGC 6156, \KS{which we identify to have} multiple \hi\ components with symmetric kinematic misalignments.  The morphology and kinematics of their \hi\ distributions suggest that these structures plausibly could be polar, a possibility that we explore with \KS{perfectly polar kinematic models that we developed to apply to the relatively low resolution and low signal-to-noise observations presented here. Although these models confirm that the anomalous gas in NGC 4632 and NGC 6156 could plausibly be polar, we cannot rule out the possibility of a strong warp (see Sec. \ref{Sec:PRGConsistency} for details); we therefore refer to these systems as \textit{potential \hi\ PRGs} throughout.}} \textcolor{black}{These systems differ from the majority of other systems with gas-rich rings \citep[e.g.][]{vanGorkom87,Serra2012,DeRijcke2013,Bait2020}, in that the hosts are gas rich late-type galaxies. }  This  expands the study of polar \hi\ structures beyond the S0 class that typically defines PRGs, but presents the additional challenge of separating the anomalous gas in the ring from that of the underlying \hi\ disk. We carry out this task in an immersive 3D virtual reality environment \citep{jarrett+21}. 
%The untargetted nature of WALLABY allows not only for such systems to be discovered in \hi, \KS{but also provides a first estimate of their incidence from a large statistical sample} given the WALLABY survey parameters and estimates of PRG detectability at different orientations and spatial resolutions. 

\KS{Recent discoveries of polar \hi\ structures are particularly interesting in the context of understanding cosmological galaxy assembly, as simulations suggest that they may result from gas accretion and infall \citep{Khoperskov2021}. This raises the possibility that the incidence of \hi-rich rings around nearby galaxies in particular -- ``\hi\ PRGs" -- could constrain this evolutionary process. While a first estimate of the incidence of \hi\ PRGs may be inferred from studies like \citet{Serra2012} 
\textcolor{black}{and \citet{Cao2022} (although in the case of \citet{Cao2022}, the misalignment is seen H$\alpha$ emission, so perhaps gaseous PRGs is more appropriate)}, the targetted nature of the existing surveys and the lack of consideration of detectability therein implies that the corresponding census is incomplete.  Obtaining a true measure of the incidence of \hi\ PRGs requires untargetted \hi\ surveys with sufficient depth, angular resolution and sky coverage to identify them, as well as a framework for assessing \hi\ PRG detectability. The observational piece of this picture will soon be in place with a new generation of widefield surveys that are underway now, such as the Widefield ASKAP L-band Legacy All-sky Blind surveY \citep[WALLABY,][]{Koribalski2020} on the Australian SKA Pathfinder \citep{Johnston2008,Hotan2021}. In this paper, we leverage the detections of potential \hi\ PRGs NGC~4632 and NGC~6156 in WALLABY PDR1 to develop the first framework for estimating \hi\ PRG detectability from untargetted \hi\ surveys, using both morphological and kinematic information and accounting for geometric and resolution effects.}

The paper is laid out as follows.  \textcolor{black}{Sec. \ref{Sec:WALLABY} examines the WALLABY \hi\ observations of NGC~4632 and NGC~6156, and Sec. \ref{Sec:Supplemental} discusses ancillary optical and mid-infrared (MIR) observations.  %\textcolor{black}{Section \ref{Sec:KinModels} presents initial attempts at kinematically modelling the \hi\ distributions of the NGC 4632 and NGC 6156 using standard methods.  However, due to some key issues in the more general models
\textcolor{black}{Section \ref{Sec:PRGConsistency} presents an exploration of whether these galaxies are consistent with simple, perfectly polar models} using a new formalism that we have developed.  Section \ref{Sec:Detectability} leverages the untargetted nature of WALLABY to estimate the incidence of \hi\ PRGs given the detectability of the NGC~4632 and NGC~6156 model polar rings determined from mocks at different ring projections and angular resolutions.}  Finally, Sec. \ref{Sec:Conclusions} gives our discussion and conclusions.  \textcolor{black}{Additionally Appendix \ref{Sec:KinModels} describes attempts at modelling warps and inclined rings, while Appendix \ref{Sec:Appendix_PRMath} describes the mathematical derivation of the perfectly polar models.}

\section{Wallaby Observations} \label{Sec:WALLABY}

%\textcolor{black}{KS FUTZING WITH THIS.}
WALLABY \citep{Koribalski2020} is an \textcolor{black}{untargetted Southern Hemisphere survey} that will detect the \hi\ gas content of $\sim 210000$ galaxies \citep{SoFiARelease} out to $z\sim0.1$. The survey will cover all of $-60^{\circ} \le  \rm{DEC} \le -15 ^{\circ}$ with some extensions reaching to $\rm{DEC} = 10^{\circ}$.  \citet{SoFiARelease} released the PDR1 observations consisting of three fields observed at the full WALLABY sensitivity centered on the Hydra and Norma clusters, and the NGC 4636 group. Each field covers 60 square degrees of the sky constructed from two adjacent tiles, which themselves are composed from two interleaved 36 beam ASKAP footprints, \textcolor{black}{providing a fairly uniform noise of $\sim1.6$ mJy/beam with a beam full-width at half-maximum (FWHM) of 30\arcsec}.  
%The actual noise levels in each field vary slightly, but are roughly $\sim1.6$ mJy/beam, and the WALLABY beam itself is $30$ \arcsec\ across.  
Once the mosaiced cubes are constructed, the \textsc{SoFiA} source finder \citep{Serra2015,Westmeier2021} is used to detect individual \hi\ sources in all fields.  The PDR1 catalogue consists of 301 \textcolor{black}{detections} in Hydra, 144 \textcolor{black}{detections} in Norma, and 147 \textcolor{black}{detections} in NGC 4636.  The majority of them are marginally resolved, with only 190 having an estimated \textcolor{black}{\hi\ distribution size along the major axis that is wider than} 4 beams. \KS{In what follows, we define the \hi\ diameter $D_{\hi}$ as the major axis extent within which the axisymmetric surface density distribution exceeds 1 $M_{\odot}~\rm{pc}^{-2}$, measured in 30" WALLABY beams.} 
% \textcolor{black}{For clarity, the radius of a galaxy, $R_{\hi}$ is defined as the point where the surface density decreases below 1 $M_{\odot}~\rm{pc}^{-2}$, and the size of a galaxy, $D_{\hi}$, is the corresponding diameter.  As noted above, the WALLABY synthesized beam has a FWHM of 30\arcsec\, so 4 beams are equivalent to 120\arcsec.}

A visual inspection of PDR1 source datacubes revealed two detections with ring-like \hi\ components that appear to have a very different geometry from that of the inner \hi\ disk: NGC~4632 (WALLABY J124232-000459), and NGC 6156 (WALLABY J163452-603705). \hi\ and optical imaging of these systems \textcolor{black}{are} presented in Figures~\ref{Fig:NGC4632_HSCOverlay}--\ref{Fig:NGC6156_DECamOverlay}, and their basic properties are listed in Table~\ref{tab:GalaxyCharacteristics}. The anomalous \hi\ components in both systems are separable from that of the main disks in position-velocity space. We do so for both galaxies using the virtual reality software package \textsc{iDaVIE}\footnote{\href{https://idavie.readthedocs.io/}{https://idavie.readthedocs.io/}} \citep{Comrie2021b}, designed to explore and interact with spectral imaging and other 3D data sets \citep{jarrett+21}. \textsc{iDaVIE}'s \textit{`paint'} tool allows direct editing of the cube masks in an immersive 3D environment, which we have used to define masks that separate the anomalous from main disk gas.  This is a particularly powerful tool for low resolution, low $S/N$ observations as it allows for adjustments at the individual pixel level while still allowing the user to view the entire cube.  However, as the separation is done via visual inspection, some quantities calculated by comparing the two components, such as the mass ratio, are approximations. 

In this section, we describe the \hi\ morphologies of NGC~4632 and NGC~6156, both as a whole and separated into main disk and anomalous components. Both here and throughout, we adopt a moment 1 colormap generated using a modified version of the \textsc{CosmosCanvas} package\footnote{\href{https://github.com/mlarichardson/CosmosCanvas}{https://github.com/mlarichardson/CosmosCanvas}}.  The colormap is designed to evoke the gas kinematics themselves, where the blueshifted gas appears to move `out of the page' and the redshifted gas appears to move `into the page'. The colormap also emphasizes differences between gas at velocities slightly above and below the systemic velocity, which is particularly useful in this study.

%  A useful analysis tool provided by 

%However, due to the anomalous gas, it was not included in the PDR1 kinematic model release (\textcolor{black}{Deg et al. 2022}). 

\subsection{NGC 4632}

  \textcolor{black}{The anomalous \hi\ gas detected by WALLABY in NGC~4632 was first reported in \citet{Lee2022}}.  
 %It is located at $\textrm{RA}=12^{\rm{h}}42^{\rm{m}}32^{\rm{s}}$, $\textrm{DEC}=-0^{\circ}4^{'}59^{''}$ and has a systemic velocity of $v_{sys}=1719~\kms$.  
 NGC~4632 is a member of a triplet system with NGC 4666 and NGC 4668 \citep{Sandage1994} and is classified as an SAc galaxy \citep{Ann2015}.  Recently \citet{Morales2018} included NGC 4632 in their search for tidal features in a subsample of the S$^{4}$G survey \citep{Sheth2010,Querejeta2015} from the \textit{Spitzer} Space Telescope imaging \citep{Werner2004}, and did not find a noticeable tidal feature in the IR imaging.  \textcolor{black}{We adopt the Cosmic Flows-III distance to NGC~4632 of 15~Mpc \citep{Tully2016}, placing it considerably more nearby than estimates from pure Hubble flow (D$\sim$24.6 Mpc, \citealt{SoFiARelease}).}  %\textcolor{black}{Given the anomalous 

\begin{figure*}
\centering
    \includegraphics[width=0.9\textwidth]{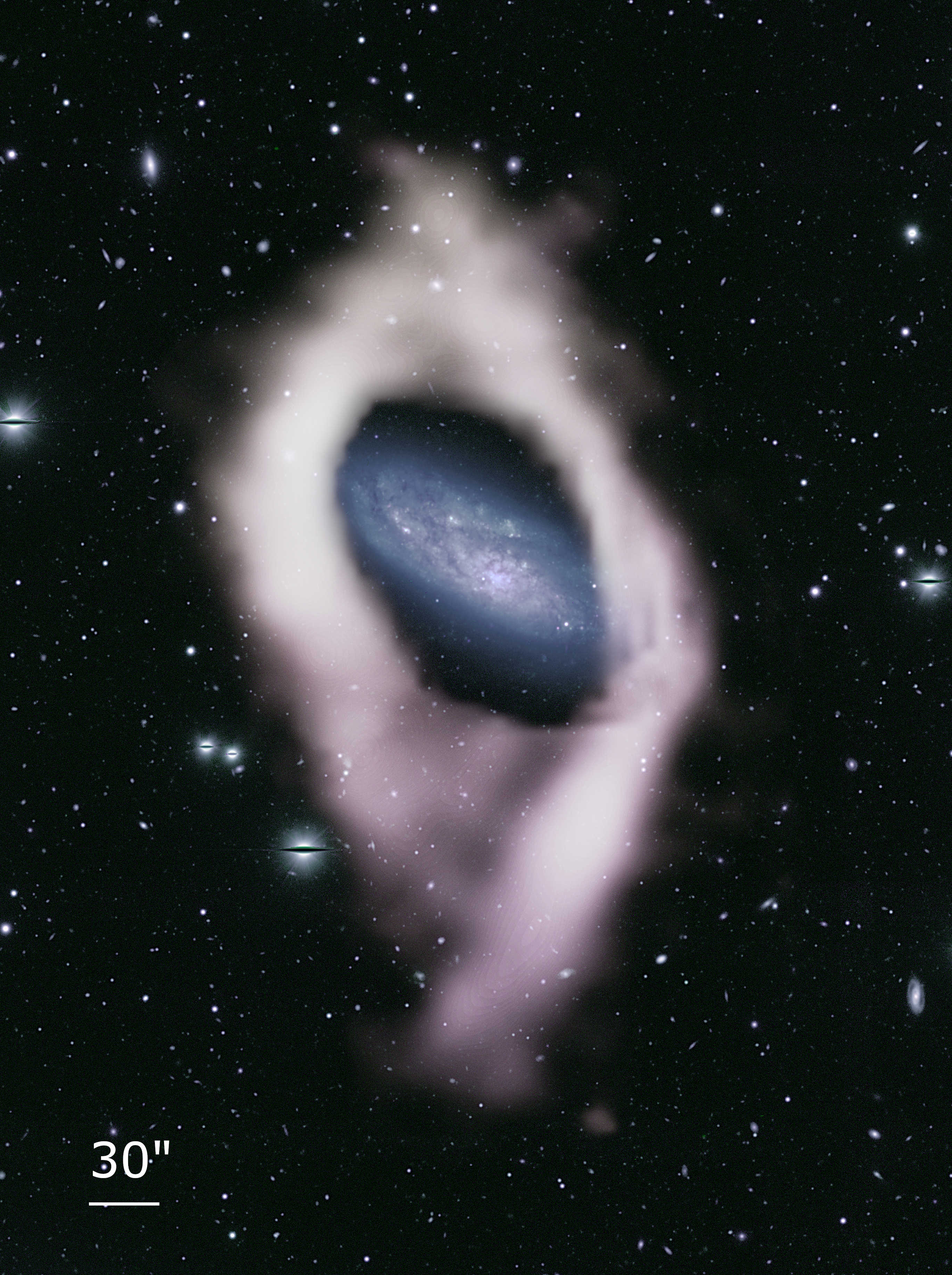} 
    \caption{\KS{Anomalous \hi\ component of the potential \hi\ PRG NGC~4632, overlayed on a composite $grz$ image from the Hyper Suprime-Cam Subaru Strategic Program (HSC-SSP, \citealt{Aihara2018}), with the white bar in the bottom-left showing the image size scale. The anomalous gas clearly has a different orientation from that of the stellar disk. The \hi\ intensity map in this figure is created by binning velocity channels together into several individual intensity maps, assigning a colour to each individual map, and blending them into one image.  This retains some velocity information while rendering the \hi\ intensity distribution, which we emphasize by using only subtle colours for velocity.  Moment 0 and moment 1 maps of the total emission, main body and anomalous \hi\ components of NGC~4632 are shown with linear intensity and velocity scales as well as with coordinate axes in Figure~\ref{Fig:NGC4632_WallabyOverview}.}}
  \label{Fig:NGC4632_HSCOverlay}
\end{figure*}

%\textcolor{black}{Jayanne to add a figure here showing the overlay of the gas and the HSC detection.}

The WALLABY detection of NGC~4632 is $\sim10$ beams in diameter, making it one of the better-resolved PDR1 sources. Figure \ref{Fig:NGC4632_HSCOverlay} shows an overlay of  \KS{the anomalous component} onto imaging taken from the Hyper Suprime-Cam Subaru Strategic Program (HSC-SSP, \citealt{Aihara2018}), \textcolor{black}{highlighting the mis-alignment of this gas relative to the underlying stellar disk}. 
%\textcolor{black}{The \hi\ intensity component of this figure is created by binning velocity channels together into several individual intensity maps and, in GIMP{\footnote{\href{https://www.gimp.org/}{GNU Image Manipulation Package}}}, assigning a colour to each individual map before blending the set of intensity maps into one image.  This retains some velocity information while rendering the \hi\ intensity distribution, which we emphasize by using only subtle colours for velocity.}

\begin{figure*}
\centering
    \includegraphics[width=0.8\textwidth]{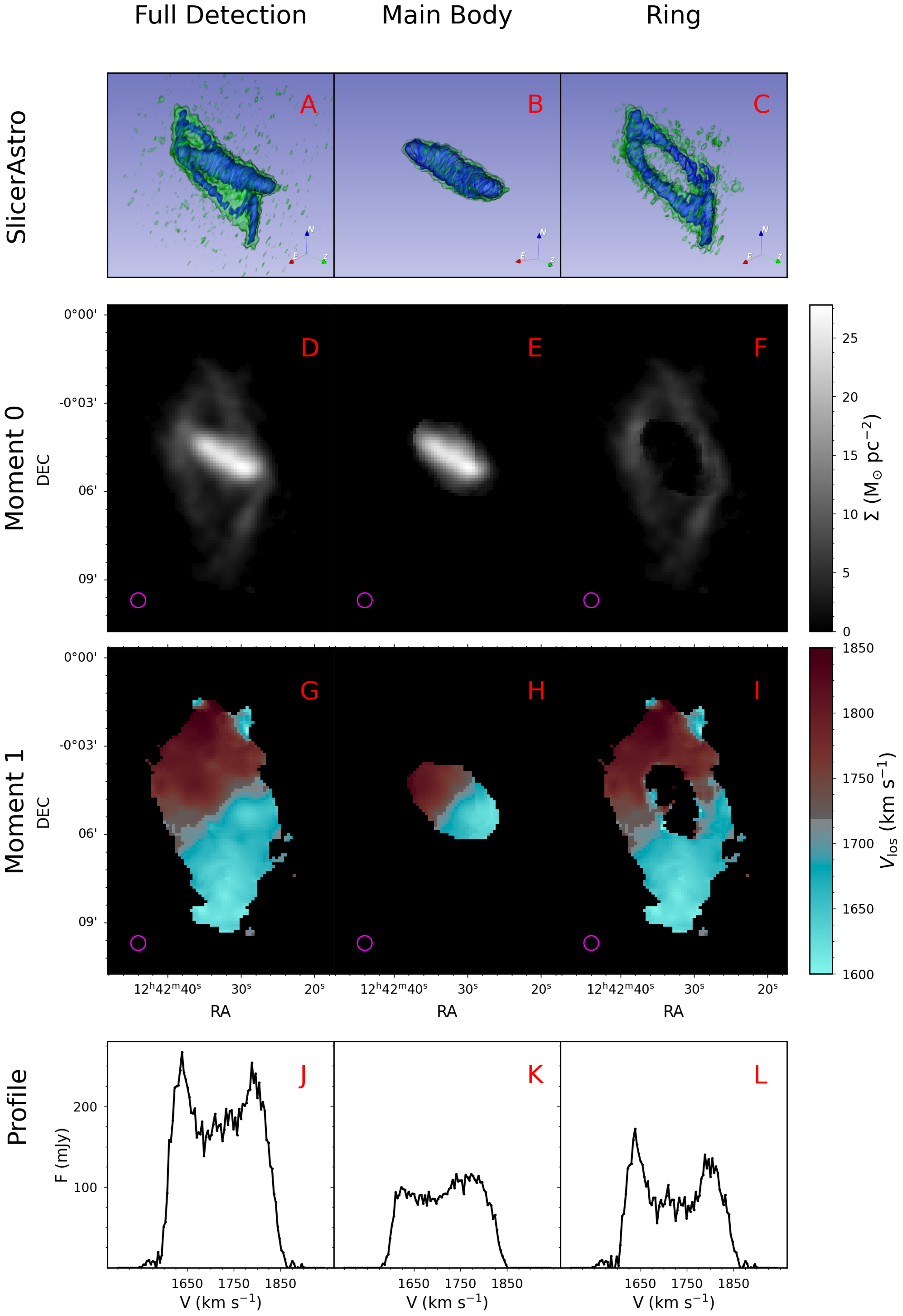} 
    \caption{\KS{An overview of the WALLABY \textcolor{black}{PDR1} \hi\ detection of NGC 4632.  From left to right, the columns show the full detection, the gas associated with the galaxy main body (found using a custom \textsc{iDaVIE} mask), and the anomalous gas.  For illustrative purposes, panels A--C in the top row show 3D projections of each component using \textsc{SlicerAstro} \citep{Punzo2017}.  In these panels, the $E,N,Z$ axes correspond to RA, DEC, and $V_{\rm{los}}$.  Panels D--F (second row) and panels G--I (third row) show moment 0 (panels D-F) and moment 1 (panels G-I) maps, respectively, with the magenta circles denoting the beam FWHM. Finally, panels J--L (bottom row) show the \hi\ profiles for each component.} 
    %gas associated with the main body of the galaxy (found using a custom \textsc{iDaVIE} mask), and the third row shows the anomalous gas.  \textcolor{black}{From left to right the columns show} an \textsc{iDaVIE} projection of the 3D gas distribution, the moment 0 map, the moment 1 map, and the \hi\ profile. \textcolor{black}{The magenta circles in the middle two columns is the beam FWHM.} The linear stretch utilized for the moment 0 maps does not highlight the low density structures that are included in the moment 1 maps, which is why the two spatial extents appear dissimilar.
    }
  \label{Fig:NGC4632_WallabyOverview}
\end{figure*}

\textcolor{black}{Figure \ref{Fig:NGC4632_WallabyOverview} shows the 3D projections, the moment 0 and moment 1 maps, and velocity profiles \KS{of the \hi\ in NGC~4632} divided into the total, main body, and anomalous gas (left, middle, and right columns).  }
%The top row of Figure \ref{Fig:NGC4632_WallabyOverview} shows a 3D projection, the moment 0 and moment 1 maps, and the \hi\ profile of the WALLABY PDR1 detection of this galaxy.  
%The 3D projection shown in this figure is generated with the  VR visualisation package. 
\textcolor{black}{In both the 3D projection and the moment 0 map,} the main body of NGC 4632 appears as a cigar-shaped structure in position-velocity space, while the anomalous gas appears as a tilted ring with a \textcolor{black}{significantly different orientation than that} of the main body.  
%Moving to the moment 0 map, the anomalous gas \textcolor{black}{appears as a} ring structure \textcolor{black}{(although some extreme warps would have a similar appearance)}.  
The ring is less distinct from the main body in the moment 1 map and velocity profile given its sky projection, an issue to which we return in Section~\ref{Sec:Detectability}. 
%approaching-receding velocity peaks that blend together due to the close spatial positions of the anomalous gas to the main galaxy body.  Finally the velocity profile shows a double peak profile characteristic of a rotating disk.  

\textcolor{black}{The mask about the main body gas is constructed using \textsc{iDaVIE} to be relatively tight in order to facilitate the kinematic modelling described in Section~\ref{Sec:PRGConsistency}.
%We use \textsc{iDaVIE} to separate the \hi\ emission into a} main body, shown in the second row of Figure \ref{Fig:NGC4632_WallabyOverview}, and anomalous gas, shown in the third row of Figure \ref{Fig:NGC4632_WallabyOverview} and highlighted in Figure \ref{Fig:NGC4632_HSCOverlay}. 
%\textcolor{black}{We adopt a relatively tight mask around the main body to facilitate the kinematic modelling described in Section~\ref{Sec:KinModels}}.
We find a ratio of main body to anomalous gas mass of $M_{\rm{anomalous}}/M_{\rm{body}}\sim 0.9$, which may be overestimated due to the tightness of the mask adopted, but \KS{the anomalous gas} is clearly a significant contribution to the total \hi\ content of NGC 4632. }
%For reasons discussed in Sec. \ref{Sec:KinModels}, we required a tight mask about the main body, which may cause some of the outlying galactic gas to be ascribed to the anomalous gas.  
%Figure \ref{Fig:NGC4632_HSCOverlay} shows an overlay of only the WALLABY anomalous gas onto imaging taken from the Hyper Suprime-Cam Subaru Strategic Program (HSC-SSP, \citealt{Aihara2018}), \textcolor{black}{highlighting the mis-alignment of this gas relative to the underlying stellar disk}.  

\begin{table*}
    \centering
    \begin{tabular}{|c|c|c|}
          &  NGC 4632 & NGC 6156\\
         \hline 
         WALLABY Name & WALLABY\_J124232-000459 & WALLABY\_J163452-603705 \\
         Type & SAc & SAB(rs)c pec \\
         RA (J2000) & $12^{\rm{h}}42^{\rm{m}}32^{\rm{s}}\pm5^{''}$ & $16^{\rm{h}}34^{\rm{m}}50^{\rm{s}}\pm13^{''}$ \\
         DEC (J2000) & $-0^{\circ}4^{'}50^{''}\pm5^{''}$  & $-60^{\circ}37^{'}2^{''}\pm9^{''}$ \\
         $V_{sys}$ ($\kms$) & $1719\pm5$ & $3266\pm15$ \\
         D (Mpc) & $15~\pm~2$  & $48~\pm~3$ \\
         \hi\ Flux (Jy km/s) & $45.5~\pm~0.5 $ & $15.3~\pm ~0.5$ \\
         $\log(M_{\hi}/M_{\odot})$  & $9.37~\pm~0.01$  & $9.92~\pm~0.02$ \\
         $\log(M_{\hi,\rm{body}}/M_{\odot})$  & $\sim9.1$  &  $\sim9.5$\\
         $\log(M_{\hi,\rm{anomalous}}/M_{\odot})$  & $\sim9.0$  & $\sim9.6$ \\
         $M_{\rm{anomalous}}/M_{\rm{body}}$ & $\sim0.9$ & $\sim1.3$\\
         $\log(M_{*}/M_{\odot})$ & $9.69\pm 0.08$ & $10.75\pm 0.08$\\
         SFR ($\Msol \textrm{yr}^{-1}$) & $0.73\pm 0.07$ & $17.60 \pm 1.83$\\
         \hline
    \end{tabular}
    \caption{\textcolor{black}{The characteristics of the two \textcolor{black}{potential \hi\ PRGs}.  The center positions and systemic velocities and associated uncertainties are from the kinematic modelling of the main body described in Sec. \ref{Sec:PRGConsistency}.  The total \hi\ flux is drawn from \citet{SoFiARelease} and the distances are from \citet{Tully2016}. The \hi\ mass is calculated using the distances and fluxes, while the stellar mass and SFR are calculated from the archival WISE detections as described in Section~\ref{Sec:Supplemental}.}  \textcolor{black}{The uncertainties in the total flux are drawn from \citet{SoFiARelease} and \KS{then propagated into $M_{HI}$. The separation into main body and anomalous \hi\ is done by eye and is rather uncertain (see Sec.~\ref{Sec:WALLABY} for details), and we therefore list approximate values for the corresponding \hi\ masses and ratio.}}}
    \label{tab:GalaxyCharacteristics}
\end{table*}

\subsection{NGC 6156}

%The second potential polar ring detection is NGC 6156 \textcolor{black}{(WALLABY J163452-603705)}.  
\textcolor{black}{The second potential \hi\ PRG, NGC~6156,} 
%is located at $\textrm{RA}=16^{\rm{h}}34^{\rm{m}}52^{\rm{s}}$, $\textrm{DEC}=-60^{\circ}37^{'}5^{''}$ with a systemic velocity of $3266~\kms$.  
\textcolor{black}{has been classified as an SAB(rs)c pec \citep{deVaucouleurs1991}, but it is important to note that it is located in the Norma cluster near the zone-of-avoidance.
%, this classification may not be fully correct.  
Despite the high extinction towards this galaxy,} NGC 6156 has been detected in IRAS \textcolor{black}{\citep{Sanders2003}} with a bright IR flux and is therefore a Luminous Infra-Red Galaxy (LIRG).

\begin{figure*}
\centering
    \includegraphics[width=\textwidth]{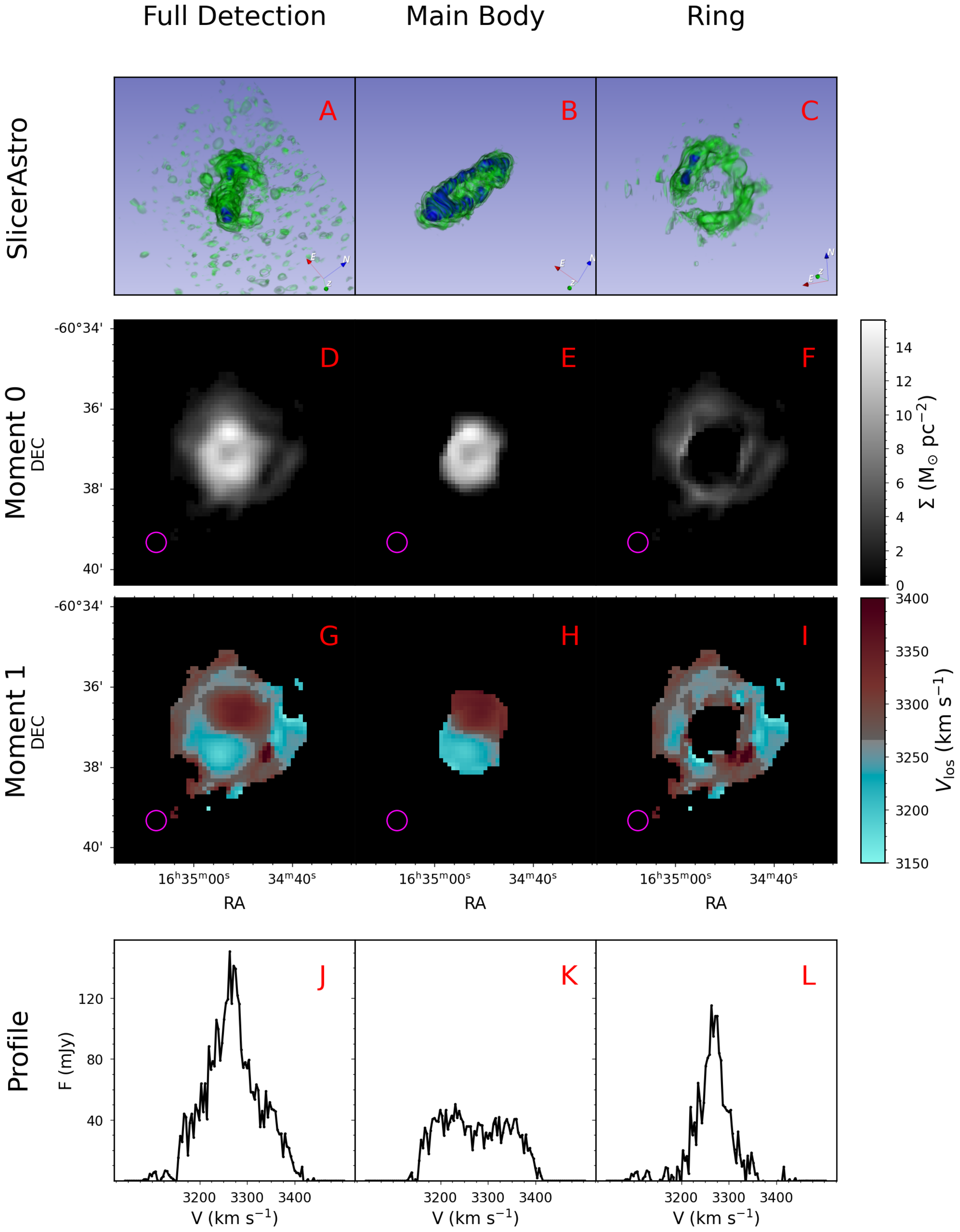} 
    \caption{An overview of \KS{the WALLABY PDR1 \hi\ detection} of NGC 6156.  The \textcolor{black}{panels are the same} as in Figure \ref{Fig:NGC4632_WallabyOverview}.}
  \label{Fig:NGC6156_WallabyOverview}
\end{figure*}

Figure \ref{Fig:NGC6156_WallabyOverview} shows an overview of the WALLABY observations of NGC 6156. In contrast to NGC~4632 in Figure~\ref{Fig:NGC4632_WallabyOverview}, the moment 0 map of the full \hi\ emission for this galaxy shows little sign of anomalous gas, whereas the 3D projection and moment 1 map show an anomalous component roughly perpendicular to the main body in position-velocity space with a bipolar velocity structure. This highlights the importance of examining the full kinematics of the gas distribution to identify anomalous components, which we discuss in the context of polar ring detectability in Section~\ref{Sec:Detectability}.  
%An examination of only the velocity profile and 2D \hi\ gas distribution would lead to a conclusion that the galaxy is simply face-on.  It is only when the velocity map, or 3D structure, is examined that the possibility of NGC 6156 being a PRG is raised.

As with NGC 4632, we use \textsc{iDaVIE} to separate out the main body from the anomalous gas in NGC~6156, and these components are shown in the \textcolor{black}{second and third columns} of Figure~\ref{Fig:NGC6156_WallabyOverview} respectively.  However, the resolution and lower $S/N$ of this detection compared to NGC~4632 makes this separation more \textcolor{black}{uncertain}. There is a projected spatial overlap between the outer main body and inner anomalous \hi\ in NGC~6156, but the anomalous gas is separated from the main body in velocity space.  Given the collisional nature of gas, it is likely separated in Cartesian space as well. It appears that the anomalous component is roughly face-on, given the low \KS{line-of-sight velocities} in the moment 1 map and single-peaked anomalous \hi\ velocity profile. 

\begin{figure}
\centering
    \includegraphics[width=0.4\textwidth]{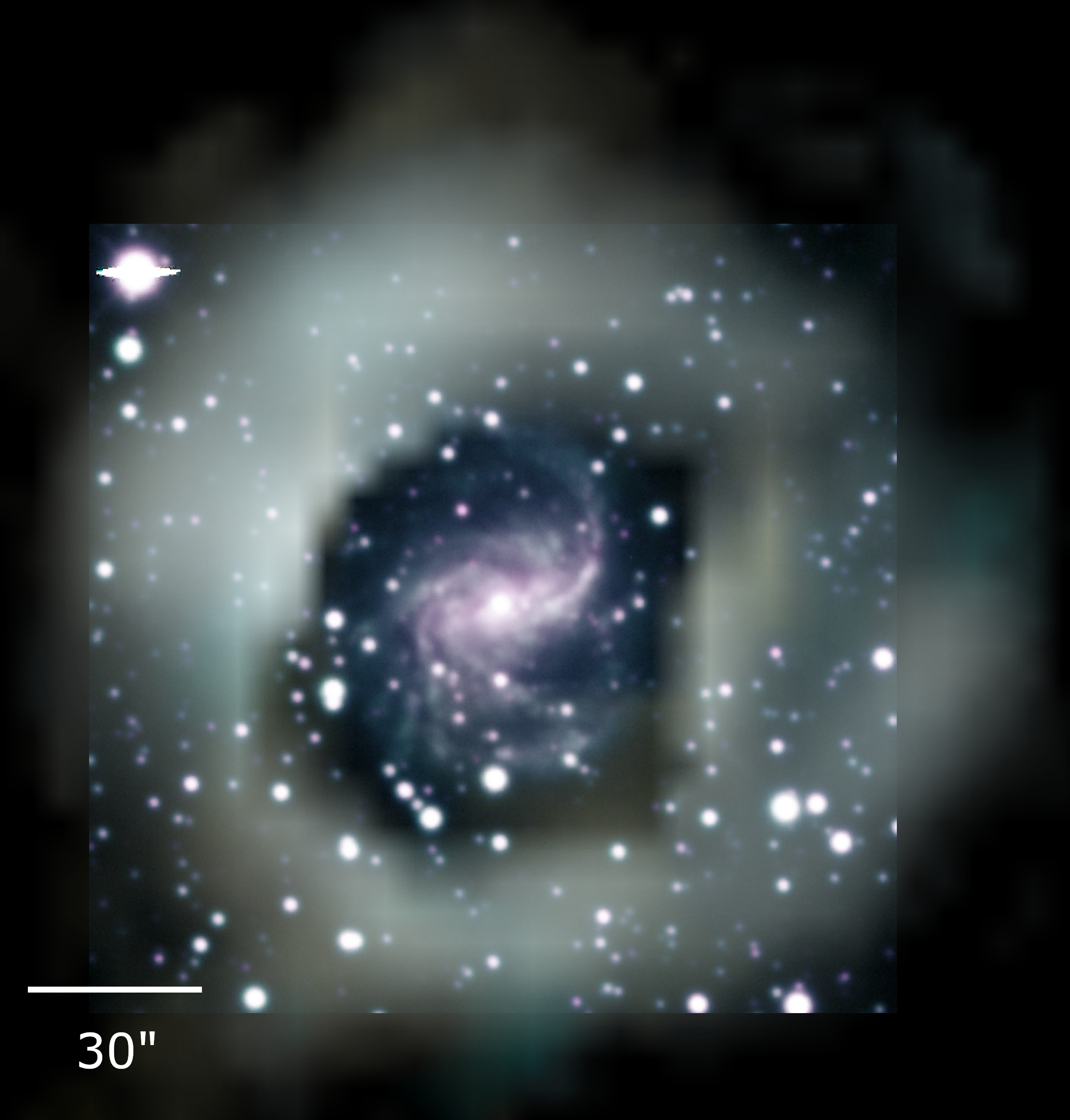} 
    \caption{\KS{Anomalous \hi\ component of the potential \hi\ PRG NGC~6156, overlayed on a composite $gr$ DECam image. The white bar in the bottom-left shows the image size scale. The \hi\ component is rendered in the same manner as in Figure~\ref{Fig:NGC4632_HSCOverlay}.  Moment 0 and moment 1 maps of the total emission, main body and anomalous \hi\ components of NGC~6156 are shown with linear intensity and velocity scales as well as with coordinate axes in Figure~\ref{Fig:NGC6156_WallabyOverview}.}}
    %Composite $g$ and $r$-band DECam image of NGC~6156 overlayed with its anomalous \hi\ component. 
    %{\textcolor{black} Depending on the reader's monitor the most redshifted gas appears subtely yellow or rust coloured while the relatively blueshifted gas appears pale cyan.}
    %in which the yellow hue denotes redshifted gas relative and the cyan hue denotes blueshifted gas.} 
    %\textcolor{black}{However, as seen in }{\textcolor{black} the intensity map and velocity field in} 
    %\color{red}Figure \ref{Fig:NGC6156_WallabyOverview}, the anomalous gas is nearly face-on and so}
    %{\textcolor{black}the velocity bins blend in this rendition, leaving few regions of pure hue.}
    %very little cyan and yellow gas is seen in this figure}.  
    %The optical cutouts do not extend as far as the anomalous \hi\ causing the box outlines seen in the image.  \textcolor{black}{The white bar shows the scale.}}
    %The anomalous \hi\ gas of NGC~6156 overlayed onto an optical image obtained from the  g and r band images of  NGC 6156 obtained from DECam observations using CTIO.  The \hi\ gas is color coded with cyan being blue shifted and yellow being redshifted.  The optical cutouts do not extend as far as the anomalous \hi\ causing the box outlines seen in the image.}
  \label{Fig:NGC6156_DECamOverlay}
\end{figure}

\textcolor{black}{Figure \ref{Fig:NGC6156_DECamOverlay} shows an overlay of the anomalous \hi\ on a composite $g$+$r$-band CTIO DECam image of NGC~6156\textcolor{black}{, created using the same approach as described %in Section~\ref{Sec:WALLABY} 
for Figure \ref{Fig:NGC4632_HSCOverlay}.} The flux ratio of anomalous to main body \hi\ is $M_{\rm{anomalous}}/M_{\rm{body}} \sim 1.3$.  As with NGC 4632, uncertainties in the mask construction \KS{render this mass ratio uncertain}, but the anomalous gas clearly contributes a great deal to the total \hi\ mass of NGC 6156.
}

\section{Supplemental Data}\label{Sec:Supplemental}

\subsection{NGC 4632}

\begin{figure}
\centering
    %\begin{subfigure}
        %\centering
        %\includegraphics[width=0.4\textwidth]{Figures/N4632AVGgrz.mir-image-2022-10-31-18-44-31.png} 
    %\end{subfigure}
    %\begin{subfigure}
        \centering
        \includegraphics[width=0.4\textwidth]{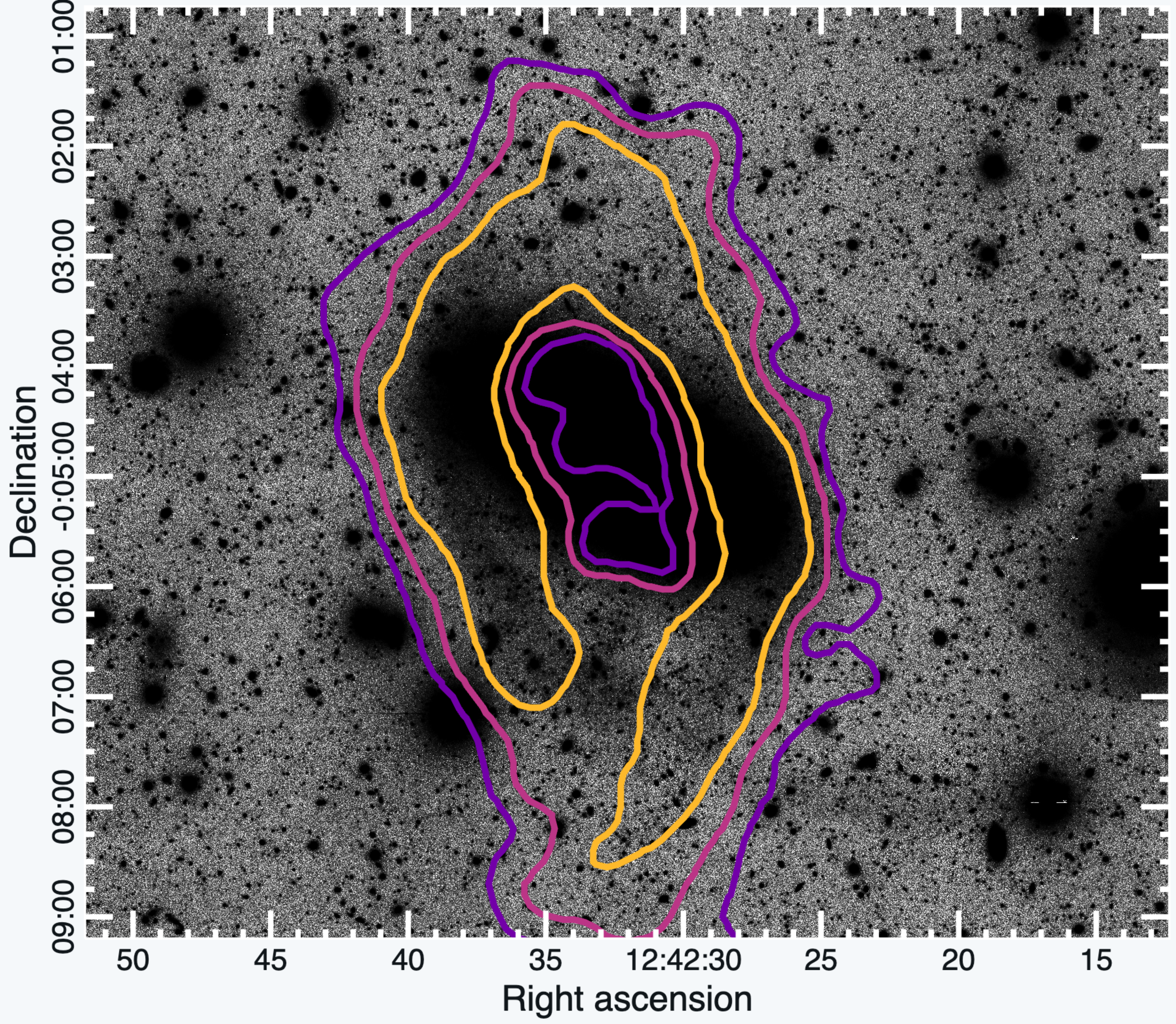} 
    %\end{subfigure}
    %\includegraphics[width=\textwidth]{Figures/N4632AVGgrz.mir-Contour-image-2022-10-31-18-40-47.png} 
    \caption{\KS{Logarithmically-stretched composite $grz$ HSC image of NGC~4632, with anomalous \hi\ component contours overlaid. The purple, pink, and orange \hi\ contours (going from dark to light) are set to 2, 6, and 12 M$_{\odot}$ pc$^{-2}$ respectively.}}
    %The HSC \textit{g}, \textit{r}, and \textit{z} stacked images of NGC 4632 overlayed with the \textcolor{black}{anomalous} \hi. The inverted grayscale image is a logarithmic stretch of the \textcolor{black}{stacked} HSC g, r, and z images adjusted using the bias and contrast in \textsc{CARTA 3.0}.  \textcolor{black}{The purple, pink, and orange \hi\ contours (going from dark to light) are set to 2, 6, and 12 M$_{\odot}$ pc$^{-2}$ respectively. }}
  \label{Fig:NGC4632_StellarRing}
\end{figure}

\textcolor{black}{The depth of the HSC imaging for NGC~4632 presented in Figure~\ref{Fig:NGC4632_HSCOverlay} affords a search for low-surface brightness optical features beyond the main disk.}  Figure \ref{Fig:NGC4632_StellarRing} shows an exposure time-weighted composite $g$, $r$, and $z$-band HSC image with a logarithmic stretch to highlight the low surface brightness features.  A clear \KS{ring-like} stellar extension is seen in the image stack that is coincident with the southern portion of the anomalous \hi\ shown by the contours in Figure \ref{Fig:NGC4632_StellarRing}, suggesting a common origin. This structure will be studied in detail in future work (Wang et al., in prep). 
\textcolor{black}{In addition to this faint feature, Figure~\ref{Fig:NGC4632_HSCOverlay} shows that the stellar disk is more extended to the Southwest than to the Northeast relative to the centre of light.}

\begin{figure*}
\centering
    \includegraphics[width=0.8\textwidth]{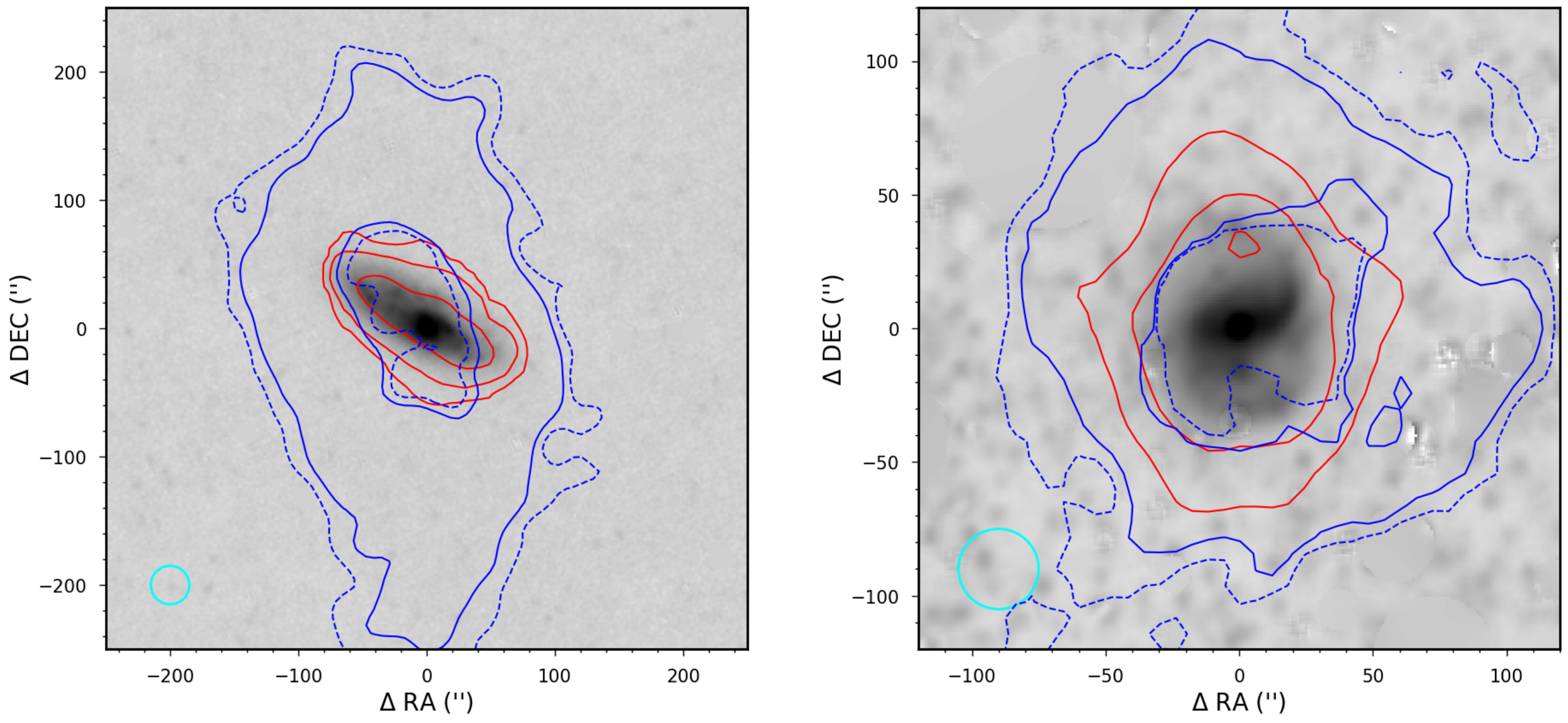} \caption{WISE W1 3.4\,$\mu$m WSXC images of NGC 4632 (left panel) and NGC 6156 (right panel) with the \hi\ gas overlayed.  \textcolor{black}{In both panels the greyscale WISE images are plotted in logarithmic units, while the main body and anomalous \hi\ are shown as red and blue contours, respectively.  For NGC 4632 in the left panel, the red contours are at $(5,10,20)$ M$_{\odot}$ pc$^{-2}$ and the dashed and solid blue contours are at $(2,6)$ M$_{\odot}$ pc$^{-2}$.  For NGC 6156 in the right panel, the red and blue contour levels are $(5,10,15)$ M$_{\odot}$ pc$^{-2}$ and $(0.5,1.5)$ M$_{\odot}$ pc$^{-2}$ respectively. \KS{The cyan circles in the bottom-left corner of both panels show the WALLABY beam FWHM, and coordinates in both panels are given relative to the centre points listed in Table~\ref{tab:GalaxyCharacteristics}.}}}
  \label{Fig:NGC4632_WISEOverview}
\end{figure*}

In addition to the WALLABY and Subaru HSC observations for NGC 4632, we obtained archival custom-built mosaics from the WISE Extended Source Catalogue \citep[WXSC;][]{jarrett+2013,jarrett+2019}.  \textcolor{black}{The left-hand panel of}
Figure \ref{Fig:NGC4632_WISEOverview} shows the WISE W1 3.4\,$\mu$m WXSC observations overlayed with the \textcolor{black}{\hi\ contours of the WALLABY moment~0 data}. \textcolor{black}{The W1 flux peak in NGC~4632 is mildly offset to the Southwest, ie.\ on the side of the disk minor axis that is more extended in the HSC imaging (Figure~\ref{Fig:NGC4632_HSCOverlay})}.

%As seen in both the WISE image and in Figure \ref{Fig:NGC4632_HSCOverlay}, NGC 4632 has a somewhat asymmetric structure.  It is likely that the origins of this stellar asymmetry and the anomalous gas are connected, but a full exploration of potential formation events is beyond the scope of this work.  

%\begin{figure}
%\centering
%    \includegraphics[width=0.4\textwidth]{Figures/NGC4632_Overlay_log_V5.pdf} \caption{The WISE W1 3.4\,$\mu$m image of NGC 4632 (grey scale) in logarithmic units with the \hi\ gas overlayed.  The blue contours show galactic \hi\ gas at contour levels of $(5,10,20)$ M$_{\odot}$ pc$^{-2}$. The dashed and solid red contours show the anomalous gas at contour levels of $(1,2.5)$ M$_{\odot}$ pc$^{-2}$ respectively.}
%  \label{Fig:NGC4632_WISEOverview}
%\end{figure}

\begin{figure}
\centering
    \includegraphics[width=0.45\textwidth]{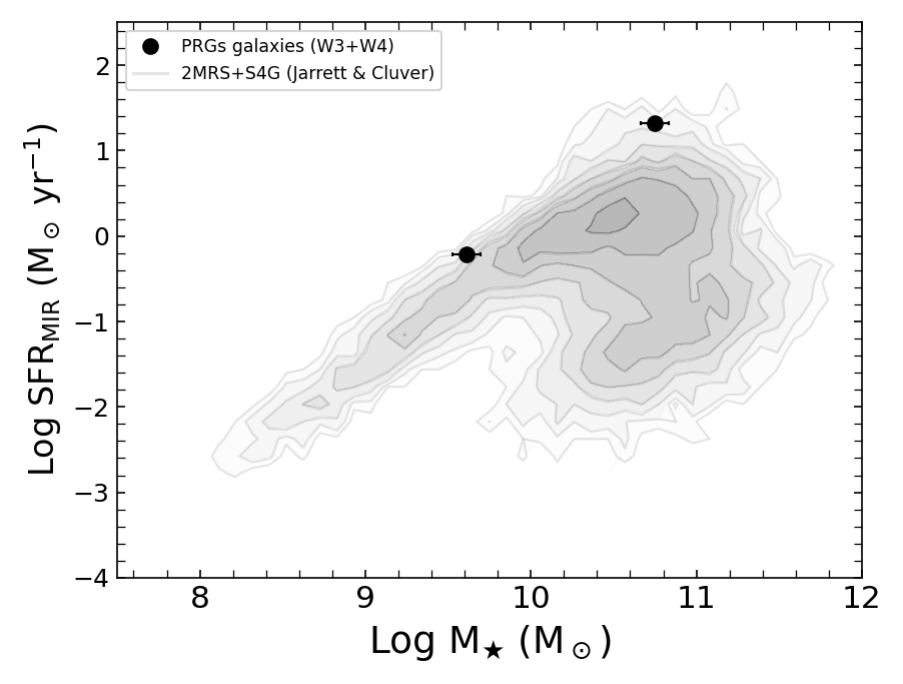} 
    \caption{The star formation rate as a function of stellar mass, both derived from WISE mid-infrared measurements.  The two \textcolor{black}{potential \hi\ PRGs} are indicated with points and estimated uncertainties, where NGC\,4632 is at lower mass and with an SFR that is consistent with normal field galaxies, and NGC\,6156 has a higher mass and excess SFR indicating a starburst phase.  For comparison purposes, the contours and greyscale represent a large sample of nearby bright galaxies from the WISE Extended Source Catalogue \citep{jarrett+2013,jarrett+2019}.
    }
  \label{Fig:SFMS}
\end{figure}

%% updated numbers from TJarrett
\textcolor{black}{We use the WISE colors and W1 luminosity \citep{Jarrett2023} to estimate the stellar mass $M_{*}$ of NGC~4632 and find $\log(M_{*}/M_{\odot})=9.69\pm 0.08$, 
making it somewhat less massive than the Milky Way.  This stellar mass coupled with the total \hi\ mass gives a gas fraction $\sim 0.3$ dex greater than the trend seen in the xGASS galaxies observed in \citet{Catinella2018}.  We also estimate the star formation rates using the W3 and W4 luminosities \citep[][see also Cluver et al. 2022, in prep]{cluver+2017} as 
$0.73\pm 0.07 ~\Msol \textrm{yr}^{-1}$,} 
%$1.13\pm 0.39 \Msol \textrm{yr}^{-1}$, 
which is well within the star formation main sequence (Figure \ref{Fig:SFMS}).

%\textcolor{black}{The WISE colors can be used to determine a number of physical properties for the galaxy. 
%WISE is able to estimate the underlying stellar mass using the colors (mapped to the stellar mass-to-light) and W1 3.4\,$\mu$m luminosity (Jarrett et al. 2022, under review).  The star formation activity is linked to the WISE W3 and W4 luminosities, as given in \citep[][; see also Cluver et al. 2022, in prep]{cluver+2017}.}
%$\log(M_{*}/M_{\odot})=9.735\pm 0.094$, 
%$\log(M_{*}/M_{\odot})=9.69\pm 0.08$, 
%making it somewhat less massive than the Milky Way.  It has a \textcolor{black}{total} star formation rate of 
%$0.73\pm 0.07 \Msol \textrm{yr}^{-1}$, 
%$1.13\pm 0.39 \Msol \textrm{yr}^{-1}$, 
%which is well within the star formation main sequence (Figure \ref{Fig:SFMS}).

\subsection{NGC 6156}

\textcolor{black}{The high extinction in the direction of NGC~6156 limits the quality of ancillary optical imaging. In the DECam imaging shown in 
Figure~\ref{Fig:NGC6156_DECamOverlay}, the spiral structure to the South appears more flocculent than that to the North.} \textcolor{black}{The WISE W1 WSXC image in Figure~\ref{Fig:NGC4632_WISEOverview} shows hints of the structure seen in the optical but with a more extended outer disk, the difference likely arising from the high optical extinction in the region. The WALLABY main body emission appears well-aligned with the W1 emission, with a similar ellipticity.} 

Following the same procedure as for NGC 4632, we use the different WISE images to estimate 
%%% updated numbers TJARRETT
 $\log(M_{*}/M_{\odot})=10.75\pm 0.08$ and SFR=$17.60\pm1.83~\Msol~\textrm{yr}^{-1}$ for NGC~6156. Again, the gas fraction, $M_{\hi}/M_{*}$, is $\sim$ 0.3 dex above the general trend of \citet{Catinella2018}, indicating that the galaxy is indeed gas-rich.  Moreover, the high star formation rate is above the star forming main sequence (see Figure \ref{Fig:SFMS}) indicating that the galaxy is currently star bursting, consistent with its bright IRAS emission \citep{Sanders2003}.

\section{Consistency with Polar Rings}\label{Sec:PRGConsistency}

\textcolor{black}{It is clear from Sec. \ref{Sec:WALLABY} that NGC 4632 and NGC 6156 contain a significant quantity of anomalous gas.  It is possible that this anomalous gas is some sort of polar structure (ring or disk), an inclined ring, or a warp.  
\textcolor{black}{Observationally and in some theoretical models,} warps and polar rings may be considered as part of the same continuum of kinematically misaligned galaxies.}
\textcolor{black}{The distinction between an extreme warp or a polar structure is ill-defined,} but in general warps exhibit a smooth transition between the main body and the anomalous gas while rings or disks exhibit a sharp discontinuity. \textcolor{black}{Kinematic modelling can distinguish between these different classes of objects, provided that the observations have sufficiently high angular resolution and $S/N$ \citep{Jorza2009, Arnaboldi1997}.} 

\KS{Unfortunately, our data are of low resolution with a correspondingly low $S/N$ (c.f.\ Figs.~\ref{Fig:NGC4632_WallabyOverview}~and~\ref{Fig:NGC6156_WallabyOverview}), making the detailed modelling required to distinguish warps from polar rings difficult. We attempt to fit warped models in Appendix \ref{Sec:KinModels} for completeness; only those applied to NGC~4632 converge to solutions, which themselves are inconclusive. In this section, we therefore adopt the more restricted approach as described below.}

\textcolor{black}{There are reasons to \KS{posit} that NGC~4642 and NGC~6156 are \hi\ PRGs.  Firstly, the anomalous gas seen in NGC 4632 appears to be well separated from the main body, strongly suggesting \KS{a ring rather than a warped disk.}  This is reinforced by the faint stellar ring in the HSC imaging is coincident with the southern portion of the anomalous gas (Fig.~\ref{Fig:NGC4632_HSCOverlay}).  Moving to NGC 6156, the anomalous gas appears to be mostly face-on, but the hint of rotation implies a position angle that is $\sim180^{\circ}$ to the main body.  Beyond these morphological clues, another reason to suspect that the anomalous gas is polar is that PRGs are far more stable and long-lived than inclined rings or extreme warps, \KS{on which the torque from the main disk are strong} \citep{Bournaud2003,Brook2008}.  As such, if the anomalous gas has an extreme morphology, \KS{it is more likely to be a polar ring than a ring or disk at an intermediate angle.} }

%\KS{Rather than exploring the full variety of warped, inclined ring, and polar ring models,   In this section, we therefore focus on building plausible models with polar rings. }

%Given the resolution and $S/N$ of these observations, it is not possible to definitively determine the morphology of the anomalous gas.  However, it is possible to explore whether the data are \textit{consistent with being a PRG}.  

\KS{We therefore consider whether or not perfectly polar models of the anomalous gas can reproduce the key features observed in the \hi\ content of NGC 4632 and NGC 6156.  We adopt a tilted-ring (TR) approach, which is widely used to kinematically model galaxies.}  \textcolor{black}{Section \ref{SubSec:TRModelling} describes TR modelling in general, while Sec. \ref{SubSec:PRGFormalism} describes our method of perfectly polar ring modelling.  Sections \ref{SubSec:NGC4632Kinematics} and \ref{SubSec:NGC6156Kinematics} present our models for NGC 4632 and NGC 6156 respectively.  }

\subsection{General Tilted Ring Modelling}\label{SubSec:TRModelling}

%TR modelling is a very successful approach for generating kinematic models of galaxies.  
The basic ideas of TR modelling were proposed in \citet{Warner1973} and \citet{Rogstad1974}, and since that time it has become one of the most widely used methods of generating kinematic models of \hi\ galaxies.  There are a variety of 2D TR modelling codes like \textsc{ROTCUR} \citep{Begeman1989,vanderHulst1992} and \textsc{2DBAT} \citep{Oh2018} that fit velocity maps (moment 1 maps).  More recently \textcolor{black}{TR modelling methods have been developed to fit 3D data cubes directly.}  Some of the more common codes are \textsc{TiRiFiC} \citep{Jorza2007}, \barolo\ \citep{diTeodoro15}, and \textsc{FAT} \citep{Kamphuis15}, but there are a number of other codes available as well (e.g. \textsc{KinMS}, \citealt{Davis2013} and \textsc{GBKFIT}, \citealt{Bekiaris2016}).

A TR model is described by a set of rings, \textcolor{black}{which in turn are characterized by} a set of geometric and kinematic parameters.  \textcolor{black}{In 3D methods, mock observations are generated and compared to observations.} These mock observations are made by first filling the rings with tracer particles and placing them in a data cube.  Then the mock cube \textcolor{black}{is} convolved with an appropriate beam \textcolor{black}{and frequency response function} so that it may be directly \textcolor{black}{compared to} the data.  The best fitting model is typically found by repeating this process and either minimizing some goodness of fit statistic or exploring the parameter space using Bayesian statistics. \KS{Below, we describe a TR  model in which the anomalous gas is a perfectly polar ring, which has many fewer free parameters than a generalized TR model in which the disk is allowed to warp \citep{Jorza2009, Arnaboldi1997}.}

\subsection{Perfectly Polar Ring Modelling}\label{SubSec:PRGFormalism}

%\textcolor{black}{General TR modelling constrains the sky plane geometry of the rings that make up the galaxy model.  
\textcolor{black}{ \KS{In a perfectly polar model, the ring} has a sky-plane inclination, \textcolor{black}{$i_{r}$}, and position angle, $PA_{r}$, such that, when rotated back to the galaxy plane using the main body inclination, $i_{g}$, and position angle, $PA_{g}$, the \textcolor{black}{ring's inclination with respect to the plane of the galaxy, $i_{r,g}$ will be} $90^{\circ}$.  \textcolor{black}{To ensure this orientation} we have developed a formalism relating \textcolor{black}{the galaxy-ring} plane geometry to the projected geometry. }

\begin{figure}
\centering
    \includegraphics[width=80mm]{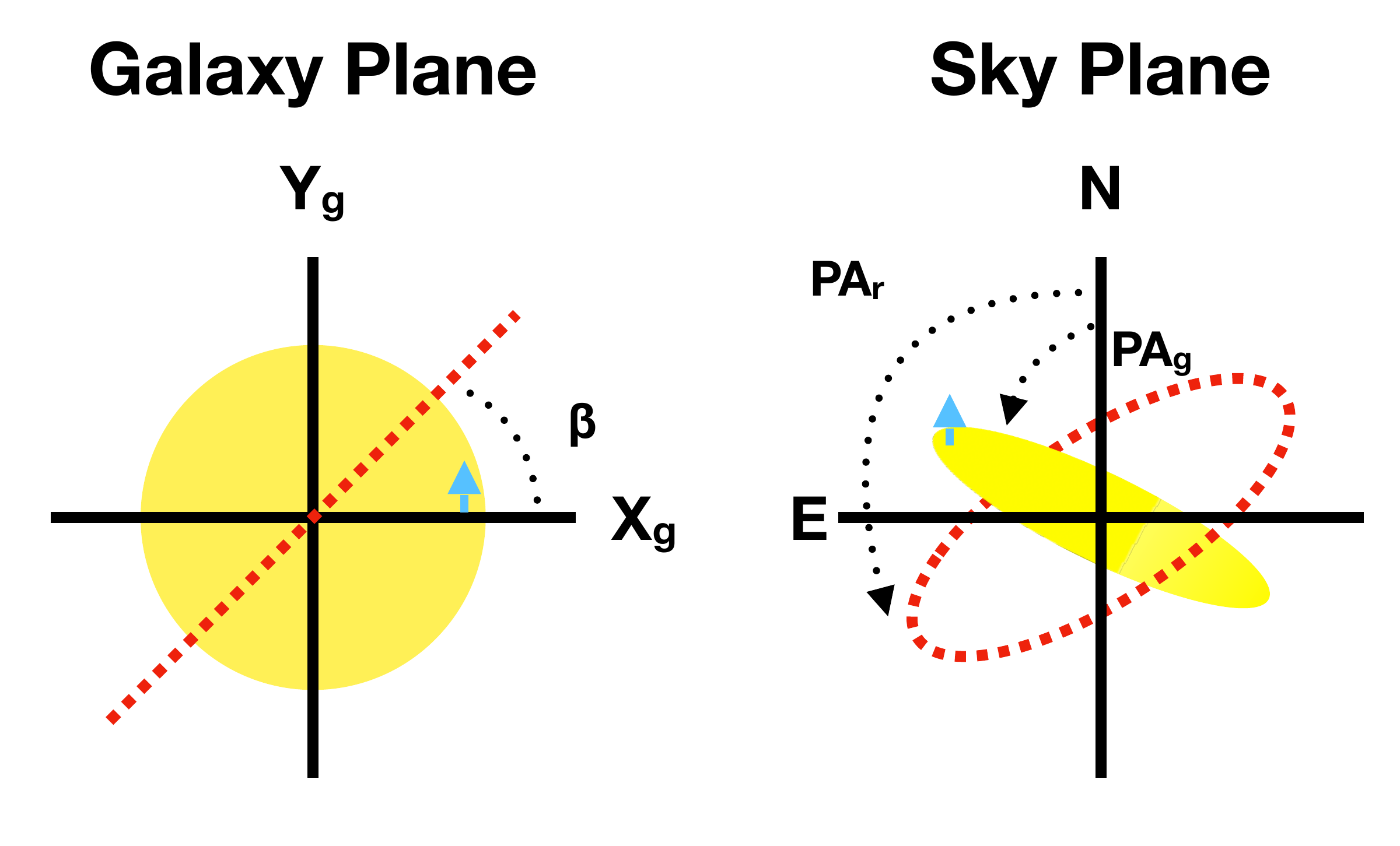} \caption{\textcolor{black}{A sketch of a PRG orientation.  The yellow circle/oval represents the host galaxy in galaxy plane coordinates (left) and sky plane coordinates (right) while the red dashed line/oval represents the polar structure.  In the galaxy plane coordinates, $(x_{g},y_{g})$, the positive $x_{g}$ axis is aligned with the approaching side of the galaxy, represented by the blue arrow in both panels.  The angle of the polar ring with respect to the $x_{g}$ axis is $\beta$, which, when coupled with the galaxy's inclination, $i_{g}$ and position angle, PA$_{g}$, uniquely determines the ring's inclination, $i_{r}$, and position angle, PA$_{r}$, in the sky plane.}
    }
  \label{Fig:RingGeometry}
\end{figure}

\textcolor{black}{The full derivation of the polar ring geometry is found in Appendix \ref{Sec:Appendix_PRMath}.  In this formalism, we have defined the angle that the ring makes with the approaching side of the galaxy as $\beta$.  This can be seen in the left-hand panel of Figure \ref{Fig:RingGeometry}, which shows a sketch of a polar ring system in galaxy plane coordinates (left) and sky plane coordinates (right). Given $\beta$ and a galaxy's observed inclination \KS{$i_g$} and position angle \KS{$PA_g$}, the ring's inclination \KS{$i_r$} and position angle \KS{$PA_r$} can be calculated.}
%Figure \ref{Fig:RingGeometry} shows a sketch of a polar ring galaxy's geometry where $\beta$ is defined as the angle of the polar ring to the approaching side of the galaxy in galaxy-plane coordinates.  As shown in the appendix, there is a relation between the main body projections and $\beta$, and a polar ring's observed inclination and position angle.  
%For notational simplicity we use the angle $\theta=PA+90^{\circ}$ rather than the position angle.  Then, when $\theta_{g}=0^{\circ}$ the inclination and position angle of a circular polar ring is
\textcolor{black}{Defining $\theta=PA+90^{\circ}$ for simplicity, and setting $\theta_{g}=0^{\circ}$, the inclination and position angle of a circular polar ring is }
\begin{equation}
    \cos(i_{r})=-\sin(i_{g})\cos(\beta)~,
    \label{Eq:IncREq}
\end{equation}
\begin{equation}
    \tan(\theta_{r})=\frac{\sin(\beta)}{\cos(\beta)\cos(i_{g})}~.
    \label{Eq:ThetaREq}
\end{equation}
Then, for any given galaxy position angle, the ring's position angle is
\begin{equation}
    PA_{r}=\theta_{r}-90^{\circ}+PA_{g}~.
\end{equation}
One thing to note is that in Eq. \ref{Eq:ThetaREq}, the $\beta$ terms have not been simplified to $\tan(\beta)$.  This is done to \KS{allow the range of $\theta_{r}$ to span $0^{\circ} \leq \theta_r \leq 360^{\circ}$}.

The relation between the observed galaxy geometry and $\beta$ to the ring's observed geometry allows for a relatively straightforward approach to kinematically \textcolor{black}{model PRGs}.  First, it is necessary to construct a `tight' mask that isolates the gas belonging to the host galaxy from the anomalous gas, \KS{and we use the \textsc{iDaVIE}  masks described in Sec. \ref{Sec:WALLABY} for this purpose. The next step is to kinematically model the galaxy body to determine its geometry and approximate outer rotation of the galaxy.}  The geometry is necessary for the polar ring modelling, and the rotation velocity can be used as a consistency check for the modelled ring.  We used the \barolo\footnote{\href{https://editeodoro.github.io/Bbarolo/}{https://editeodoro.github.io/Bbarolo/}} code \citep{diTeodoro15} to generate the galaxy model as it is fast and can be run with a user supplied mask.  This ensures that the model is \KS{only applied to} the gas contained within the `tight' mask.  Given the low resolution and $S/N$, during this step the galaxy body is modelled as a `flat' disk with a constant inclination and position angle.  Following \citet{Deg2022}, we fit 2 rings per beam.  Finally, we model the anomalous gas as a perfectly polar ring.  For simplicity we treat the gas as single ring with three free parameters, $\beta$, $v_{rot}$, and $\Sigma$ (the ring surface density).  This is a limited enough number of free parameters that they can be explored using a basic grid search.  To be clear, this step involves comparing the combined ring plus galaxy model to the \textsc{SoFiA}-masked datacube.

%\begin{enumerate}
%    \item Construct a tight mask that isolates the galaxy.
%    \item Model the masked galaxy as a flat disk (constant geometric parameters across the galaxy) using a TR algorithm.  For this work, %we have used the \textsc{3DBarolo} code \citep{diTeodoro15} as it is relatively fast and can be run with a user defined mask easily.
%    \item Model the ring with three free parameters, $\beta$, $v_{rot}$, and $\Sigma$ (the ring surface density).  Given the limited %number of free parameters, this can be done using a basic grid search.
%\end{enumerate}

%%In this work, we have used \textsc{iDaVIE} to construct the `tight' masks about the main body of the galaxy.  These are seen in middle rows of Figs. \ref{Fig:NGC4632_WallabyOverview} and \ref{Fig:NGC6156_WallabyOverview}.  For both cases, a comparison of the middle (galaxy body) and bottom (anomalous gas) rows \textcolor{black}{shows} that the masking has isolated the main bodies of the galaxies quite well.  In particular, for NGC 6156, the apparently Gaussian profile has been transformed into a more standard, albeit noisy, flat/double-peaked profile.  

\subsection{NGC 4632}\label{SubSec:NGC4632Kinematics}

Armed with the `tight' mask for NGC 4632, we are able to proceed with the modelling.  \textcolor{black}{An additional difficulty in the TR modelling of the galaxy is the importance of the initial inclination estimate.  In \barolo\ the code will either estimate the TR geometric parameters (position angle and inclination) from the observed moment maps or the user can supply estimates.  In low resolution, low $S/N$ observations, \barolo\ tends to find final models with very similar parameters to those initial estimates due to the flatness of the likelihood surface.  In fact, we find that \barolo\ is able to produce acceptable fits for initial estimates of $i_{g}\ge55^{\circ}$.}  The obvious choice is to select the optical inclination as the initial estimate ($\sim65^{\circ}$ based on the Subaru HSC images).  But, a secondary problem arrives at high inclinations. \textcolor{black}{As the galaxy becomes more edge-on, Eq. \ref{Eq:ThetaREq} shows that the difference between the galactic position angle and the ring position angle increases}.  \textcolor{black}{Supplying \barolo\ with an initial estimate of $i_{g}=65{^\circ}$ yields a final model with $i_{g}=67^{\circ}$.  Such an inclination for the galaxy body causes ring models to have too large of a difference in position angles between the galaxy and the ring to match the observations.}  In order to obtain a position angle difference similar to that seen in the upper right panel of Figure \ref{Fig:NGC4632_WallabyOverview}, \textcolor{black}{we chose an initial estimate of $i_{g}=60^{\circ}$ for \barolo, which yields a final galaxy body model with $i_{g}=62.5^{\circ}$.}

%For the fit, we set the model resolution at 2 rings/beam, which is fairly standard for 3D TR modelling \citep{Deg2022}.  The galaxy body is set to be a 'flat' disk with constant geometric parameters across all \textcolor{black}{rings.  The} `tight' mask is used in order to force \barolo to fit only the galaxy body rather than the polar ring.  With these constraints and the initial inclination estimate, \barolo is able to produce a reasonable galaxy body model with $i_{g}=62.5^{\circ}$.  

\begin{figure*}
\centering
\includegraphics[width=\textwidth]{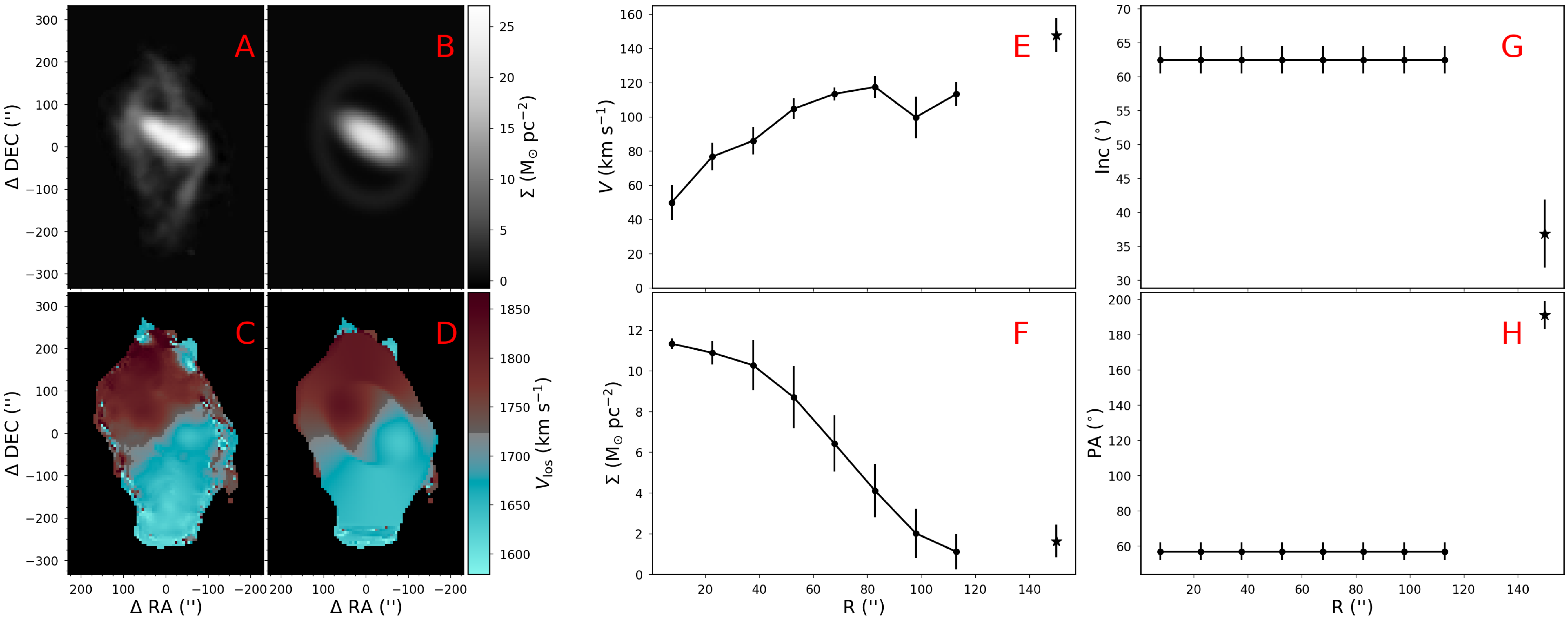} \caption{The best fitting \KS{perfectly polar ring model} of NGC 4632.  
The moment 0 maps of the galaxy and the model are shown in panels A and B respectively, while the moment 1 maps are shown in panels C and D. The velocity scale in panels C and D is centered at the systemic velocity of the warped \barolo\ model.  The rotation curve, deprojected surface density, inclination, and position angle profiles are shown in panels E-H respectively.  The polar ring parameters are indicated by stars and are the last radial point in each of the right-hand panels. As with Figs. \ref{Fig:NGC4632_WallabyOverview} and \ref{Fig:NGC6156_WallabyOverview}, the moment 0 maps use a linear stretch which explains the differing spatial extents of the moment 0 and moment 1 maps.}
  \label{Fig:NGC4632Model}
\end{figure*}

With the galaxy body modelled, it is then possible to model the polar ring using a different set of software than \barolo.  \textsc{MCGSuite}\footnote{\href{https://github.com/CIRADA-Tools/MCGSuite}{https://github.com/CIRADA-Tools/MCGSuite}} (Mock Cube Generator Suite; \citealt{Lewis2019}, Spekkens et al. in prep) is a code that generates standalone TR models from either scaling relations or directly from a set of ring parameters.  We model the polar ring by searching the relevant TR parameters \textcolor{black}{and} using \textsc{MCGSuite} to produce mock data cubes that can be compared to the observed data.  To get the orientation, rotation speed, and surface density, a grid search of $\beta$, $v_{rot}$, and $\Sigma$ is performed.  The \textsc{MCGSuite} realization of each full model (body + ring) is then compared to the \textsc{SoFiA}-masked data.  Figure \ref{Fig:NGC4632Model} shows the final best fitting model\textcolor{black}{, as determined by having the lowest $\chi^{2}$ value during the mock cube - observation comparison}.  \textcolor{black}{This model has $\beta=335^{\circ}\pm 5^{\circ}$.}

\KS{Figure \ref{Fig:NGC4632Model} illustrates that the best-fitting model reproduces some of the key features in the NGC~4632 \hi\ detection: the velocity map shows the dual peaks seen in the data.  In addition, the difference between the main body position angle and the ring position angle broadly agrees with the observed structure. Nonetheless, there are also differences between the model and the data, such as the sharpness of the transition between the galaxy body and ring region and the inclinations of the main body and ring. It is possible that these differences imply that the ring that is not perfectly polar, but our attempt a modelling such as structure is not a better fit to the data on the whole (see Appendix~\ref{Sec:KinModels}). Given that that the perfectly polar model reproduces key features of the observations (a ring like structure with a similar velocity map), it is reasonable to conclude that NGC 4632 is indeed plausibly an \hi\ PRG.}

\subsection{NGC 6156}\label{SubSec:NGC6156Kinematics}

NGC 6156 is analyzed using the same procedure as NGC 4632.  The `tight' body mask is input into \barolo and the galaxy body is modelled.  As with NGC 4632, supplying an initial estimate for the inclination is critical.  The optical observations seen in Fig. \ref{Fig:NGC6156_DECamOverlay} as well as the images of NGC 6156 obtained in the Carnagie-Irvine Galaxy Survey (CGS, \citealt{Ho2011}), \KS{suggest that the galaxy ($\sim40^{\circ}$).  This is consistent with aspect ratio of the \hi\ moment 0 map in Fig. \ref{Fig:NGC6156_WallabyOverview}E, but inconsistent with the degree of rotation seen in the moment 0 map of Fig. \ref{Fig:NGC6156_WallabyOverview}H.  The WISE image in Figure \ref{Fig:NGC4632_WISEOverview}, which we expect to more reliably trace the disk geometry in this high-extinction region, has a morphology that appears more inclined than the optical image and more consistent with the main body \hi\ velocity.}

\KS{Given the moment 1 map of the galaxy main body and the morphology of the anomalous gas as well as the optical morphology, we } provide \barolo\ with an initial 
 inclination estimate of $i_g=50^{\circ}$, yielding a galaxy model with $i_g = 51^{\circ}$.  NGC 6156 is an even lower resolution and $S/N$ detection than NGC 4632.  Thus the small shift from the initial inclination estimate is unsurprising given flatness of the likelihood surface. \KS{Given the question of the inclination, we did attempt to model the galaxy using a $40^{\circ}$ initial estimate for the main body.  However, this model yielded a ring that is even more inclined, with lower rotation velocity, in poorer agreement with the data than the model adopted below.}

\begin{figure*}
\centering
    \includegraphics[width=\textwidth]{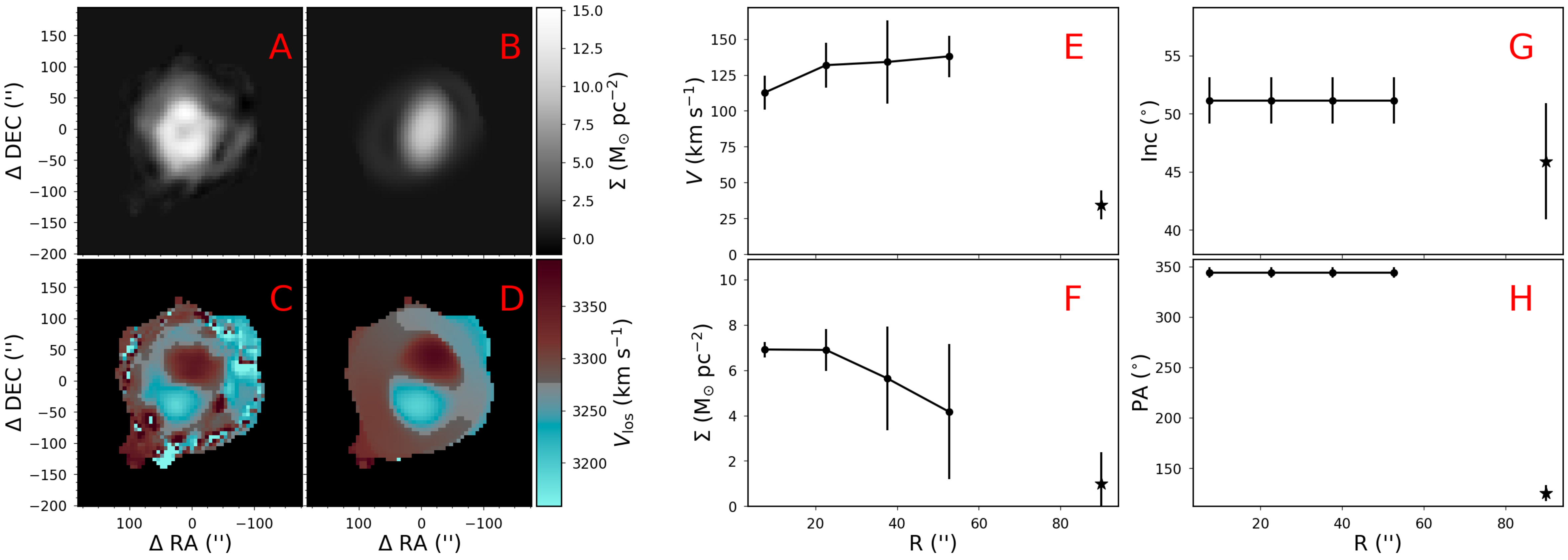} 
    \caption{The best fitting \KS{perfectly polar ring} model of NGC 6156.  The panels are the same as in Figure \ref{Fig:NGC4632Model}. }
  \label{Fig:NGC6156Model}
\end{figure*}

With the main body model fit from \barolo the same grid-based search of $\beta$, $v_{rot}$, and $\Sigma$ can be done for the ring.  As with NGC 4632, 3D realizations of each body+ring model are made using \textsc{MCGSuite} and the \textsc{SoFiA}-masked observed data cube is compared with these realizations to find the best fit.  Figure \ref{Fig:NGC6156Model} shows the best fitting model for NGC 6156\textcolor{black}{, which has $\beta= 153^{\circ} \pm 5^{\circ}$}.

\KS{Figure \ref{Fig:NGC6156Model} shows that the best-fitting perfectly polar model for NGC~6156 produces a moment 1 map with a similar velocity structure to the data. However, model moment 0 map has a ring that is more distinct, more inclined, and at a somewhat different position angle from the observed gas. Moreover, the low ring velocity relative to the outermost point of the galaxy's rotation curve raises the question of whether or not it is dynamically stable. These may be indications that the anomalous gas is better described by a warped disk or a ring with $i_{r,g} \ne 90^{\circ}$ than a polar ring. However, the warped disk models that we attempted to apply to these data did not converge to a solution (see Appendix~\ref{Sec:KinModels}), precluding further investigations in this regard until higher-resolution and $S/N$ data are available. Thus while the perfectly polar models of NGC~6156 do exhibit differences from the data, it remains possible that the anomalous gas is a polar structure and and alternate model could be not be found. We therefore consider NGC~6156 to be a plausible \hi\ PRG given the quality of the available data.}

\section{Polar Ring Detectability in WALLABY}\label{Sec:Detectability}

The detection of two \textcolor{black}{potential \hi\ PRGs} in the WALLABY pilot fields is exciting and challenging as they were unexpected.  \citet{Reshetnikov2011} estimated the incidence rate of \textcolor{black}{PRGs}, $f_{i}$, in the local Universe to be $\sim 1/1000$. \textcolor{black}{If the two are indeed \hi\ PRGs, then given the $\sim200$ PDR1 detections that are resolved by more than 4 beams is certainly inconsistent with the canonical rate.} \textcolor{black}{However, hints of a higher incidence rate for \hi\ PRGs are seen in \citet{Serra2012}, as well as in the ionized gas study of \citet{Cao2022}.  These both show incidence rates closer to $1\%$, but they are also both targetted surveys, which makes an inference of the true incidence rate much more difficult due to their selection functions.  As an untargtetted survey, future WALLABY data releases will certainly contain enough detections to determine if there is indeed a higher incidence rate of \hi\ PRGs.}

\textcolor{black}{Given the untargeted observing approach for WALLABY}, the incidence rate of \textcolor{black}{\hi\ PRGs} in  PDR1 can be estimated \textcolor{black}{from the two \textcolor{black}{potential \hi\ PRGs} that were identified} using a relatively simple process.  Firstly, detecting a PRG must depend on the geometry of the system and the resolution of the galaxies.  That is, there is a geometric detection rate $f_{g}(b)$ where $b$ is the resolution of the galaxy \textcolor{black}{in beams}.  \textcolor{black}{Then} the number of detected \textcolor{black}{PRGs} should be
\begin{equation}\label{Eq:FullDetectionRate}
    N_{\rm{detected}}=f_{i} \int f_{g}(b) N(b) db~,
\end{equation}
where $N_{\rm{detected}}$ is the number of detected \textcolor{black}{\hi\ PRGs}, $b$ is the resolution of the galaxies, and $N(b)$ is the number of galaxies at a given resolution.  \textcolor{black}{We note that if NGC~4632 and NGC~6156 harbour strong warps instead of true \hi\ PRGs, the incidence rate we estimate nonetheless applies to these extreme kinematically distinct components.}

In order to estimate $f_i$, it is first necessary to quantify $f_{g}(b)$. To understand the dependence of \hi\ PRG detectability on geometry and resolution, we generated a suite of mock PRG cubes based on our best fitting NGC 4632 and NGC 6156 models using \textsc{MCGSuite}.  \textcolor{black}{The suite consists of 49 cubes for each model at resolutions of 3, 4, 6, 8, and 10 beams across for a total of 490 mock observations.  These cubes span a range of main body inclinations between $0^{\circ}$ and $90^{\circ}$ and $\beta$ values between 0$^\circ$ and 180$^\circ$.}  Each cube is generated with a noise of 1.6 mJy/beam, matching them to WALLABY.  In order to mimic a WALLABY analysis, \textsc{SoFiA-2} is run on the cubelets to generate a mask.  \textcolor{black}{We then visually examined each moment 0 and moment 1 map individually to determine if a signature of a polar ring is clearly apparent.  If the feature can possibly be mistaken for noise, we do not count it in the determination of $f_{g}(b)$.  This determination is subjective, so the calculation of $f_{g}(b)$ from these maps should be considered as an approximation.}

\begin{figure*}
\centering
    \includegraphics[width=\textwidth]{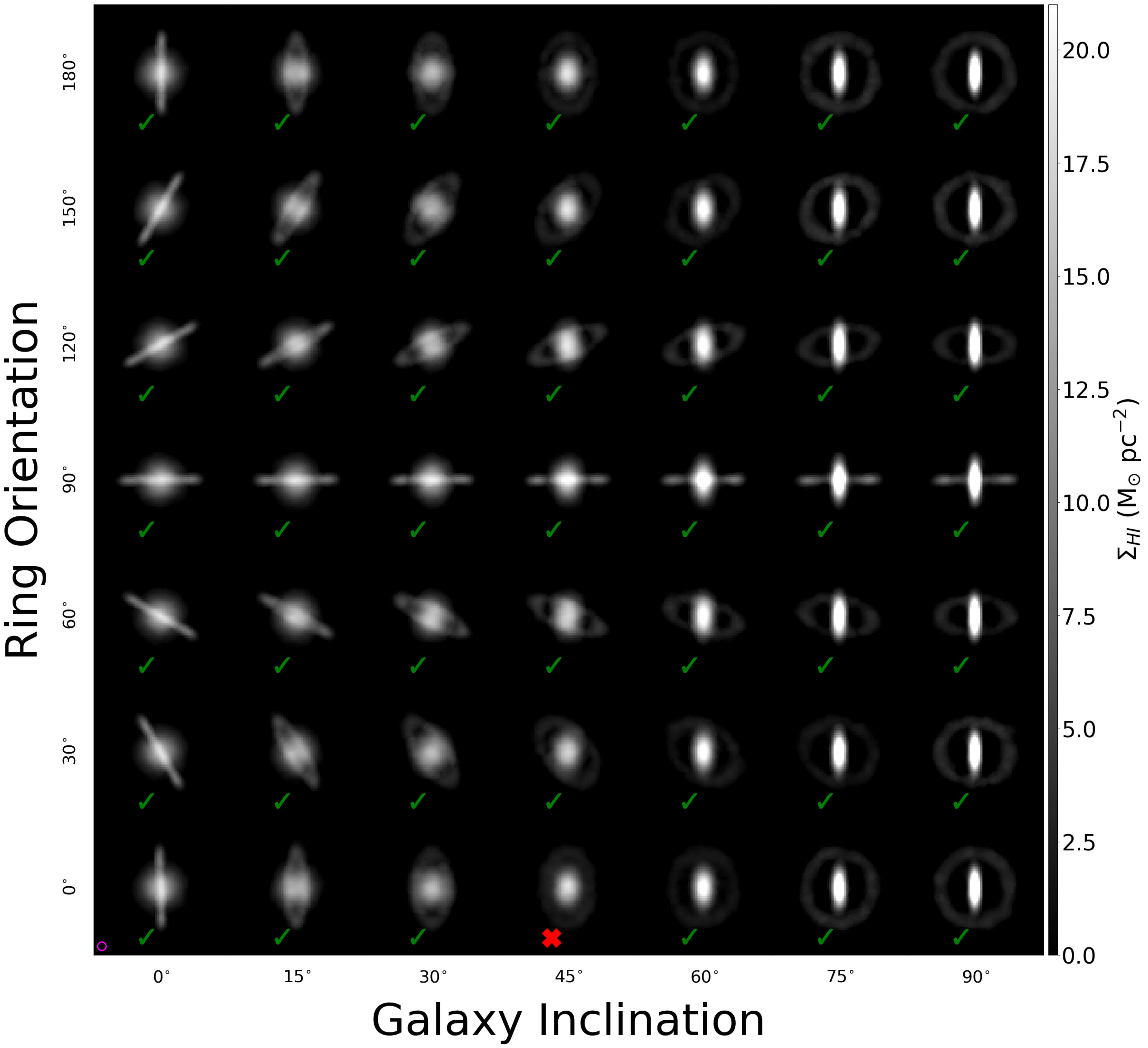} \caption{Moment 0 maps \textcolor{black}{computed from different projections of the best fitting \hi\ PRG model} for NGC 4632, where the observed geometric parameters \textcolor{black}{$\beta$ (y-axis)} and $i_{g}$ (x-axis) are varied.  These models are built to \textcolor{black}{have resolutions} of 10 beams across, which is consistent with observations of NGC 4632.  The small magenta \textcolor{black}{circle represents} the beam size. The noise level in each model cube is 1.6 mJy/beam, which is consistent with WALLABY.  The green checkmarks and show models that are \textcolor{black}{recognizable} as \textcolor{black}{PRGs} from the moment 0 map\textcolor{black}{, whereas the red x denotes the model that would not be recognizable.}}
  \label{Fig:NGC4632_Mom0_10BA}
\end{figure*}

\begin{figure*}
\centering
    \includegraphics[width=\textwidth]{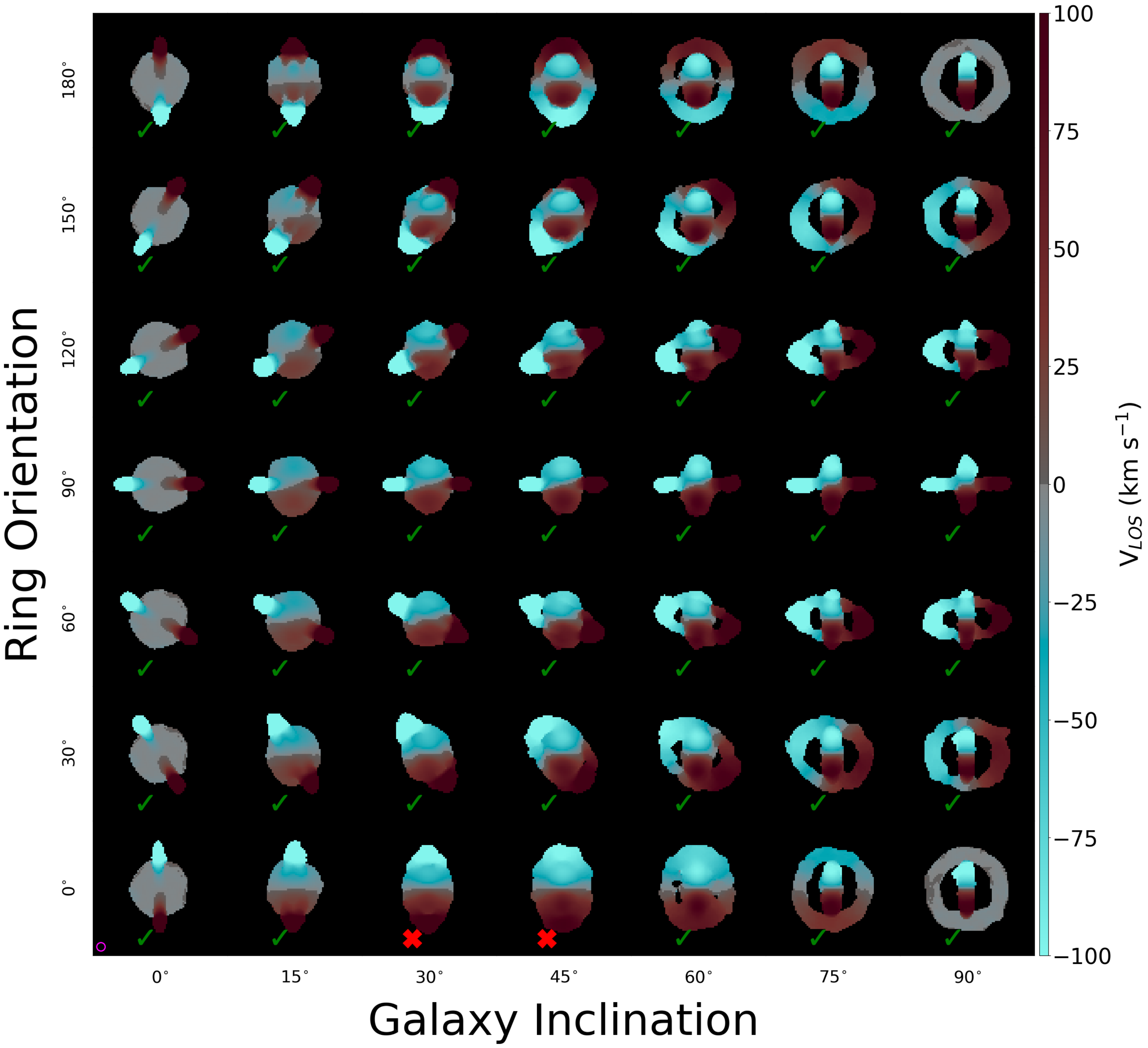} \caption{\textcolor{black}{Same as Figure \ref{Fig:NGC4632_Mom0_10BA}, \textcolor{black}{but for the moment 1 maps of different projections of the best fitting \hi\ PRG model} of NGC 4632 at a 10 beam resolution.}
    %Moment 1 maps using the best underlying model for NGC 4632 where the observed geometric parameters \textcolor{black}{$\beta$ (y-axis)} and $i_{g}$ (x-axis) are varied.  These models are built to \textcolor{black}{have resolutions} of 10 beams across, which is consistent with observations of NGC 4632.  The small magenta \textcolor{black}{circle represents} the beam size. The noise level in each model cube is 1.6 mJy/beam, which is consistent with WALLABY. The green checkmarks and red x's show models that are \textcolor{black}{recognizable} as \textcolor{black}{PRGs} from the moment 1 map. }
    }
  \label{Fig:NGC4632_Mom1_10BA}
\end{figure*}

Figures \ref{Fig:NGC4632_Mom0_10BA} and \ref{Fig:NGC4632_Mom1_10BA} show the moment 0 and moment 1 maps for the 10 beam resolution NGC 4632 models.  \textcolor{black}{The} moment 0 panels are shown using the same brightness scale for all panels to show the difference that the inclination makes in the detectability.  Similarly, a linear stretch is used for the brightness as this is the typical stretch used when examining many galaxies in a large survey to find interesting outliers, like \textcolor{black}{PRGs}.  At this resolution, which is roughly the observed resolution of NGC 4632, all orientations in the moment 0 maps show a clear ring structure except perhaps for the $\beta=0^{\circ}$, $i_{g}=45^{\circ}$ panel.  For the moment 1 maps in Figure \ref{Fig:NGC4632_Mom1_10BA}, the situation is similar, with most orientations showing a clear sign of a polar ring.  However, in the $\beta=0^{\circ}$, $i_{g}=30^{\circ}$ and $45^{\circ}$ panels, the velocity structure of the polar ring and main disk are orientated such that the polar ring signature is hidden. 

\begin{figure*}
\centering
    \includegraphics[width=\textwidth]{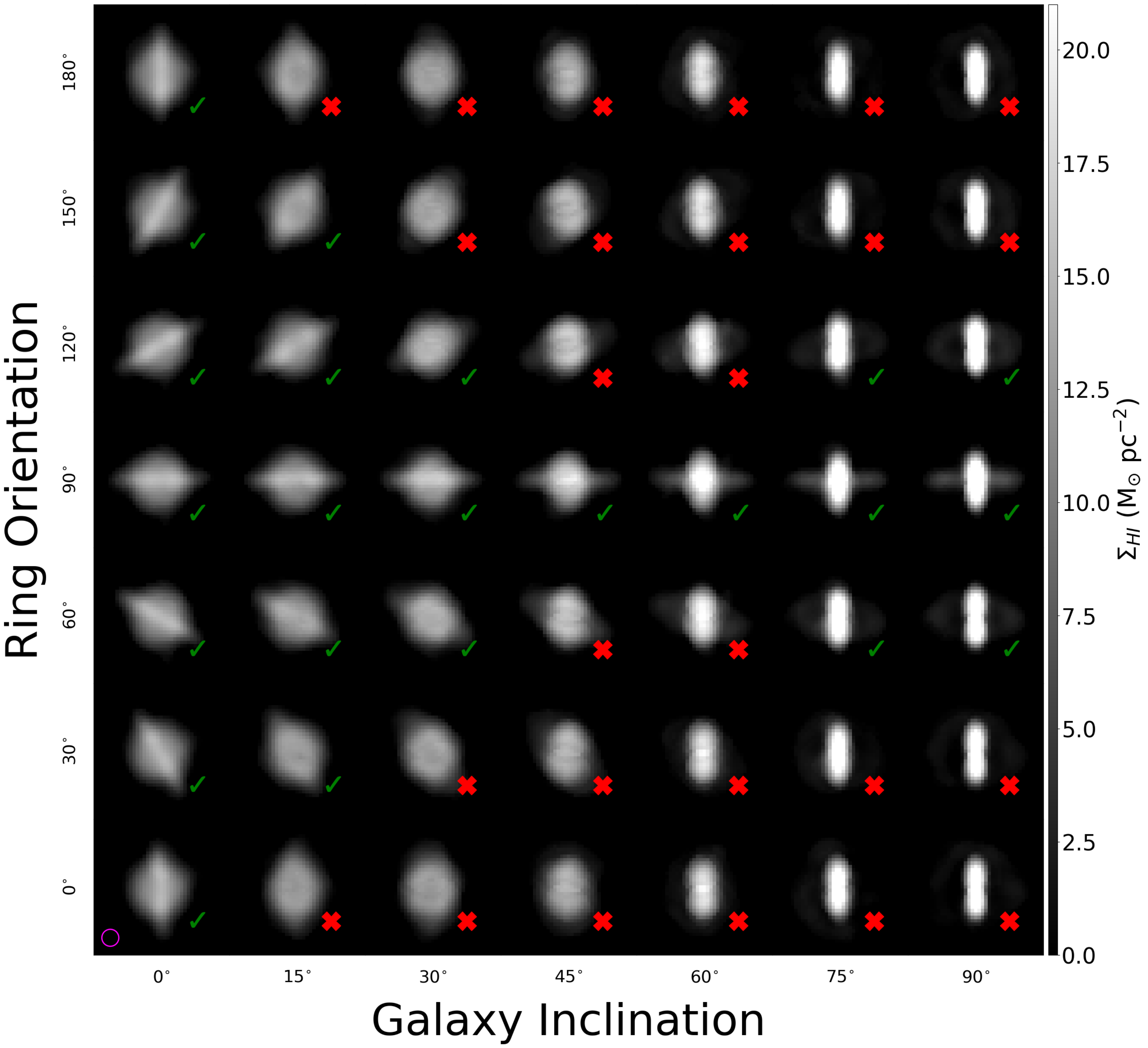} \caption{Same as Figure \ref{Fig:NGC4632_Mom0_10BA}, but \textcolor{black}{but for the moment 0 maps of different projections of the best fitting \hi\ PRG model} of NGC 6156 at a 6 beam resolution.
    %Moment 0 maps using the best underlying model for NGC 6156 where the observed geometric parameters \textcolor{black}{$\beta$ (y-axis)} and $i_{g}$ (x-axis) are varied.  These models are built to \textcolor{black}{have resolutions} of \textcolor{black}{6 beams across}, which is consistent with observations of NGC 6156.  The small magenta \textcolor{black}{circle represents} the beam size. The noise level in each model cube is 1.6 mJy/beam, which is consistent with WALLABY.  The green checkmarks and red x's show models that are \textcolor{black}{recognizable} as \textcolor{black}{PRGs} from the moment 0 map.
    }
  \label{Fig:NGC6156_Mom0_10BA}
\end{figure*}

\begin{figure*}
\centering
    \includegraphics[width=\textwidth]{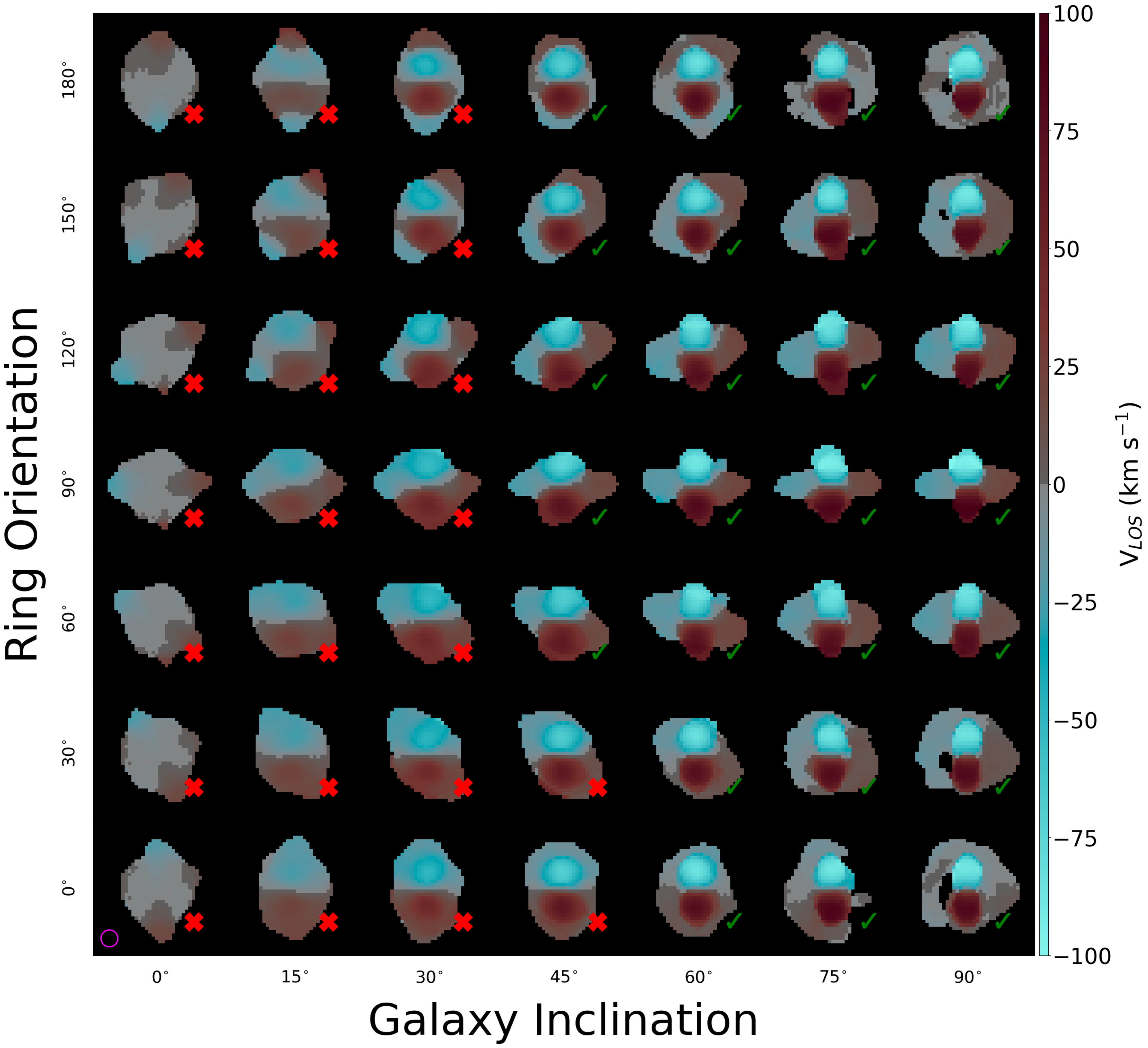} \caption{Same as Figure \ref{Fig:NGC4632_Mom0_10BA}, \textcolor{black}{but for the moment 1 maps of different projections of the best fitting \hi\ PRG model} of NGC 6156 at a 6 beam resolution.
    %Moment 1 maps using the best underlying model for NGC 6156 where the observed geometric parameters \textcolor{black}{$\beta$ (y-axis)} and $i_{g}$ (x-axis) are varied.  These models are built to \textcolor{black}{have resolutions} of \textcolor{black}{6 beams across}, which is consistent with observations of NGC 6156.  The small magenta \textcolor{black}{circle represents} the beam size. The noise level in each model cube is 1.6 mJy/beam, which is consistent with WALLABY.  The green checkmarks and red x's show models that are \textcolor{black}{recognizable} as \textcolor{black}{PRGs} from the moment 1 map.
    }
  \label{Fig:NGC6156_Mom1_10BA}
\end{figure*}

A more complicated picture emerges when looking at the NGC 6156 models at 6 beams across shown in Figs. \ref{Fig:NGC6156_Mom0_10BA} and \ref{Fig:NGC6156_Mom1_10BA}.  While the ring is clear in the low inclination panels of Fig. \ref{Fig:NGC6156_Mom0_10BA}, as well as the \textcolor{black}{$\beta=90^{\circ}$} panels, at intermediate \textcolor{black}{inclinations the} low and high $\beta$ orientations do not indicate the presence of a polar ring. It is worth noting here that the PRG candidate shown in Figure 1 of \citet{Nishimura2022} bears a striking resemblance to the $i_{g}=0^{\circ}$ moment 0 maps.  Moving to the moment 1 maps, a clear polar ring signature is seen in most of the high inclination models.  The presence of a rotating disk surrounded by low rotation gas seen in many of the panels is a typical polar ring signature that may not always be appreciated when examining velocity maps.  It is in fact this signature that suggested that NGC 6156 is a PRG initially.

An interesting point to note about the 6 beam resolution models is that the moment 0 and moment 1 maps are complementary.  At intermediate resolutions, the orientations where the ring is most easily detected in the moment 0 maps are where the velocity maps provide little indication of a ring structure.  Similarly, where the moment 1 maps easily see polar rings are the orientations where the moment 0 maps provide less information.  This complementarity may allow WALLABY and other interferometric studies to detect many more \textcolor{black}{PRGs} than previous works (provided \textcolor{black}{the observations} are of sufficient resolution).

\begin{figure*}
\centering
    \includegraphics[width=\textwidth]{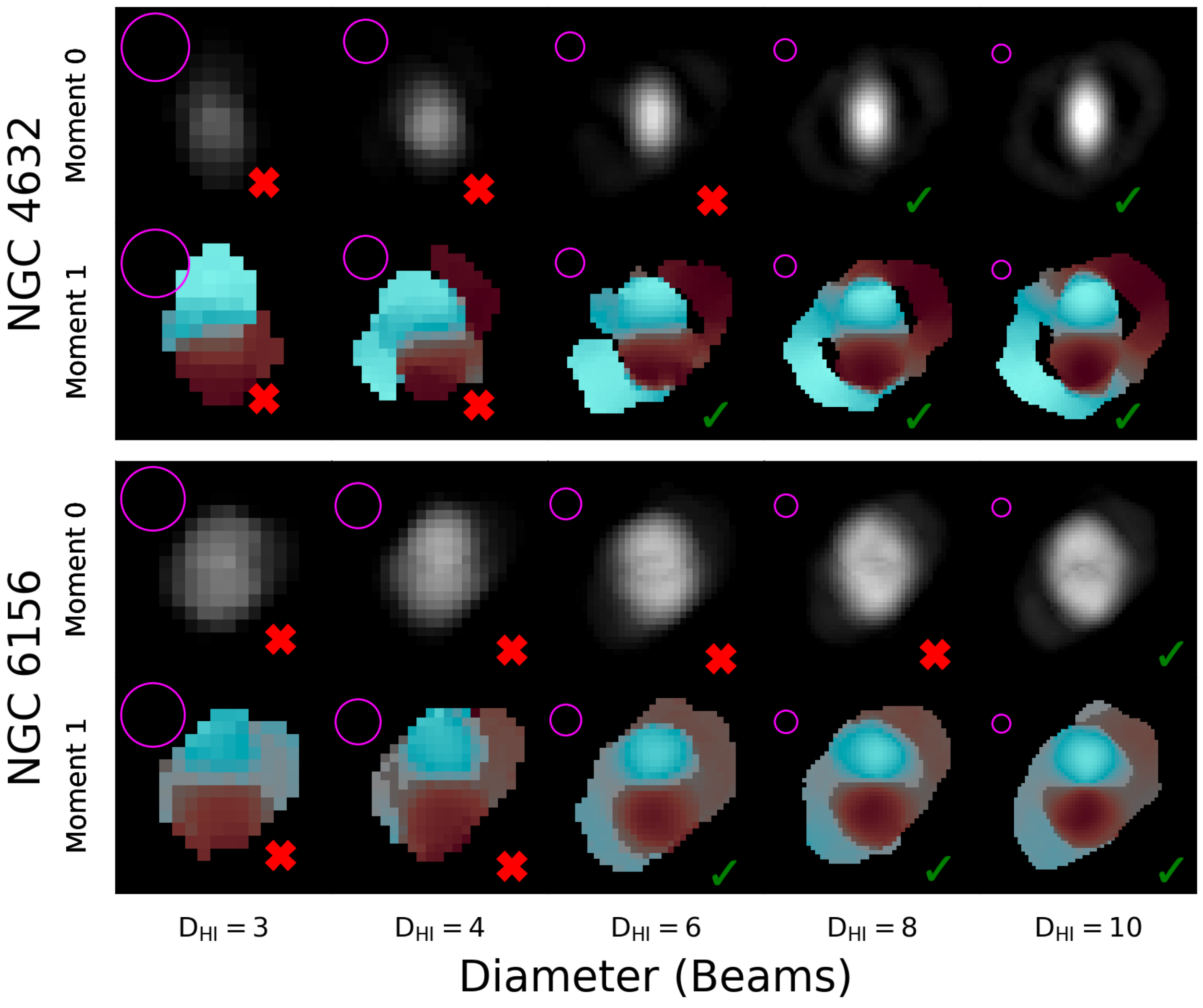} \caption{A demonstration of how angular resolution affects the detectability of \textcolor{black}{\hi\ PRGs} \textcolor{black}{from both moment 0 and moment 1 maps}. This figure shows both galaxies at a single orientation ($i = 45^\circ$, $\beta = 30^\circ$) at the five resolutions that were examined. The circles at the top left corner of each frame represents the beam size for that figure. The green checkmarks and red x's indicate panels where we were able or unable to determine the galaxy is PRG based on the specific moment map respectively.}
  \label{Fig:resfig}
\end{figure*}

While the NGC 4632 and 6156 models are significantly different, a key reason for the lower detectability in Figs. \ref{Fig:NGC6156_Mom0_10BA}-\ref{Fig:NGC6156_Mom1_10BA} compared to Figs. \ref{Fig:NGC4632_Mom0_10BA}-\ref{Fig:NGC4632_Mom1_10BA} is the resolution.  To highlight this effect, Figure \ref{Fig:resfig} shows the moment maps for a single geometry of the NGC 4632 model (top rows) and NGC 6156 model (bottom rows) at different resolutions.  In both cases, at 3 beams across no polar features can be detected and at 10 beams across the polar ring is clearly detected.  

%Mock observations were conducted at 49 orientations and 5 resolutions. Each galaxy was modelled at 3 Beams Across (BA), 4 BA, 6 BA, 8 BA, and 10 BA. The actual observation resolution for NGC 4632 was 10 BA, while the actual observation resolution for NGC 6156 was 6 BA. These 10 figures, each containing 49 plots, were then examined to determine how many orientations yielded a detectable PRG model. From these 490 plots, the detectability fraction of PRGs may be extrapolated for each resolution.

\begin{figure*}
\centering
    \includegraphics[width=\textwidth]{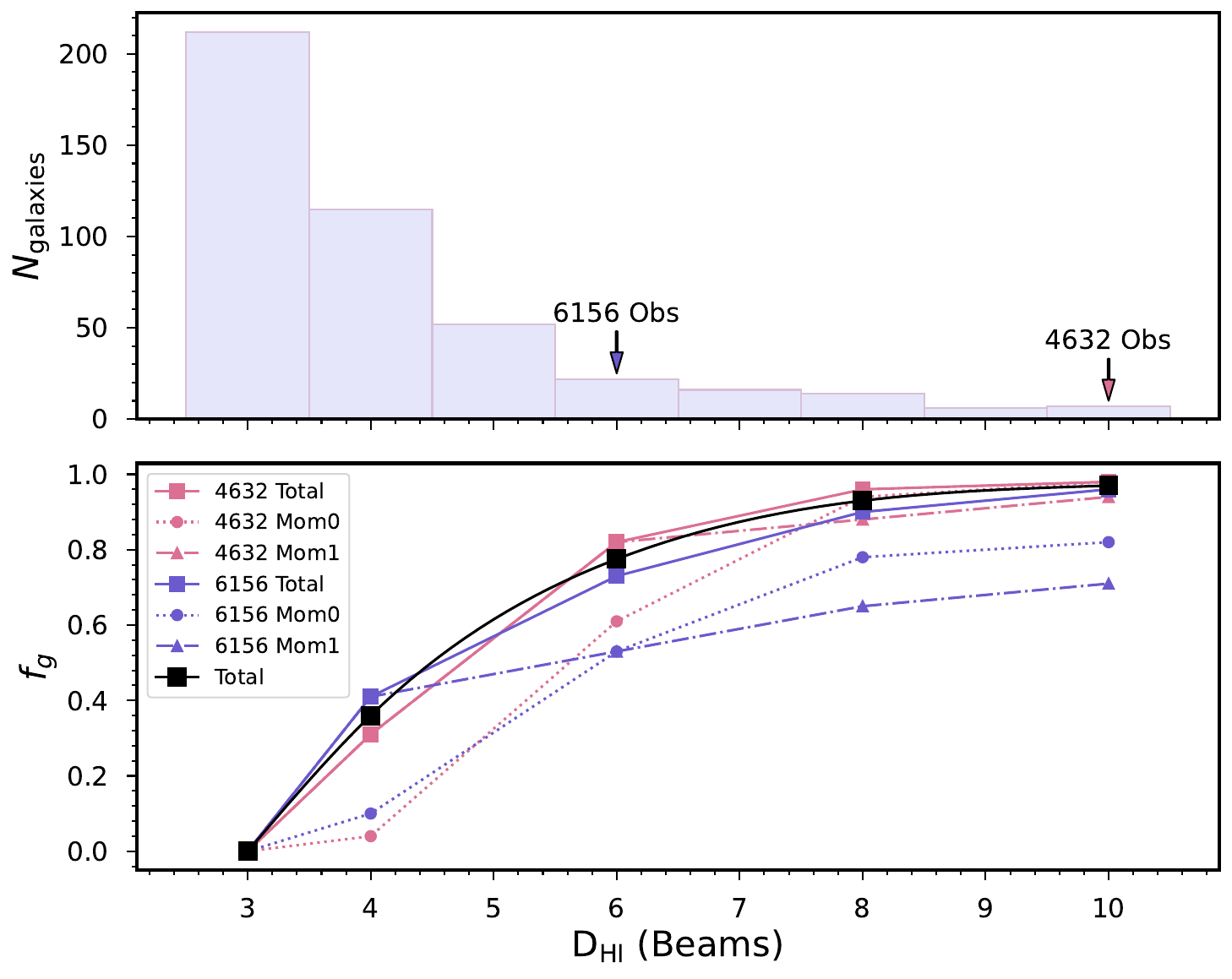} \caption{The number of observed PDR1 galaxies as a function of \textcolor{black}{the number of beams across the major axis of that detection} (top panel), and the \textcolor{black}{estimated fraction of \hi\ PRGs that would be detectable at that resolution} (bottom panel).  In the top panel, the the arrows indicate the resolution of each the \textcolor{black}{detected potential \hi\ PRGs}. \textcolor{black}{In the bottom panel, the dotted, dashed and solid lines represent the detectability fraction from moment 0, moment 1 and the combination of moment 0 and moment 1 information respectively for the NGC~4632 model (magenta) and the NGC~6156 model (blue).}}
  \label{Fig:probplot}
\end{figure*}

\textcolor{black}{For the purposes of quantifying $f_{g}(b)$ using the full suite of models, if a polar ring feature is identifiable in either the moment 0 or moment 1 map of a particular model, then we consider that PRG model detectable.} The bottom row of \textcolor{black}{Figure \ref{Fig:probplot}} shows $f_{g}(b)$ for the NGC 4632 model (red lines) and NGC 6156 model (blue lines).  An interesting trend seen is that, at low resolutions, polar ring features are more easily detected in the moment 1 maps, but at high resolutions they are more easily seen in the moment 0 maps.  In both models, at 10 beams across $f_{g}\sim1$, while at 3 beams across $f_{g}\sim0$.

To obtain a total detectability fraction at a given resolution, the NGC 4632 and NGC 6156 detectability fractions are averaged together. And to apply this at any resolution, we applied a spline fit to the points.  The full curve for $f_{g}(b)$ is shown as the black line in the bottom row of \textcolor{black}{Figure \ref{Fig:probplot}}.

Armed with a $f_{g}(b)$, it is possible to estimate the incidence of \textcolor{black}{PRGs} in WALLABY PDR1.  Rearranging and discretizing Eq. \ref{Eq:FullDetectionRate} gives
\begin{equation}\label{Eq:FullIncidenceRate}
    f_{i}=\frac{ N_{\rm{detected}} } {\sum_{b} f_{g}(b) N(b)}.
\end{equation}
The upper row of \textcolor{black}{Figure \ref{Fig:probplot}} shows $N(b)$ \textcolor{black}{of the WALLABY PDR1 detections}, while the bottom row shows $f_{g}(b)$, allowing an incidence rate to be calculated relatively easily.  Using $N_{\rm{detected}}=2$, $f_{i}\approx 2 \%$, which implies that there are approximately 8 \textcolor{black}{PRGs} among the 592 galaxies in WALLABY PDR1, despite only 2 being detected.  In order to estimate the uncertainty on $f_{i}$, we repeated the calculation using $N_{\rm{detected}}=1$ and $N_{\rm{detected}}=3$, yielding $f_{i,\rm{low}}\approx 1\%$ and $f_{i,\rm{high}}\approx 3\%$ respectively.  

\textcolor{black}{This detection rate is a simple approximation based on a number of assumptions that may prove incorrect.  The biggest source of error is whether our \textcolor{black}{potential \hi\ PRGs} are in fact extreme warps.  This is perhaps unlikely for NGC 4632, which would require a fairly extreme warp to reproduce the morphology of the \hi\, but it is certainly possible for NGC 6156.  Additionally, our geometric detection analysis is based on visual classifications that are certainly subjective.  Moreover, we have not explored the effects of confusion due to the warps that are present in nature.  Given this, our high detection rate should be taken as an upper limit on the occurrence of PRGs in WALLABY PDR1.}

\section{Conclusions}\label{Sec:Conclusions}

Using \hi\ alone, we have detected 2 \textcolor{black}{potential \hi\ PRGs} in the WALLABY pilot fields, NGC 4632 and NGC 6156.  NGC 4632 shows a clear ring structure in the moment 0 map, while NGC 6156 shows a characteristic ring pattern in the velocity map.  

\textcolor{black}{Once the existence of the anomalous gas was recognized in the moment maps, we investigated the plausibility of this gas being consistent with a polar ring. To that end, we} first separated that anomalous gas from the host galaxy's gas using the \textsc{iDaVIE} VR interactive tool.  This allowed for the construction of kinematic models for the host galaxies using \barolo.  We then developed a formalism relating the projected inclination and position angle of a polar ring to the host galaxy's inclination, position angle, and the angle the ring makes with respect to the approaching side of the galaxy.  This allowed the exploration of perfectly polar models to find reasonable fits to the data using \textsc{MCGSuite}.

The kinematic model fit for NGC 4632 is significantly better than the fit for NGC 6156, but it is not perfect.  The model moment 0 map contains more of the features seen in the observations compared to the NGC 6156 model and \textcolor{black}{observations}.  \textcolor{black}{However, our perfectly polar models are very strict and do not encompass the full range of PRG's seen in nature.  Polar structures may be slightly elliptical and have inclinations from the galaxy body in the range $75^{\circ}\ge i_{r,g}\ge90^{\circ}$.  Exploring these parameters will likely yield much better fits to the data and comparing those models to those of extreme warps will help to confirm whether NGC 4632 is indeed a PRG. At this point, we argue that NGC 4632 is a \textcolor{black}{potential \hi\ PRG} until such sophisticated modelling and comparisons to deeper observations are completed.}
%\textcolor{black}{Nathan, you should expand this into another sentence, or at least add some quantification to this statement}.
%Future analysis considering elliptical rings as well as deviations from polarity will provide models for the data, and allow for a deeper understanding of different ring formation scenarios.  Nonetheless, these relaxations from our perfectly polar models are still within the general definition of \textcolor{black}{PRGs}, and it is \textcolor{black}{likely} that NGC 4632 is indeed a PRG.

For the second galaxy in our study, the model fit for NGC\,6156 represents a bigger challenge.   The lower resolution and $S/N$ of NGC 6156 causes the modelling and interpretation of the anomalous gas to be more ambiguous.
%much more difficult. 
The likely orientation of the ring structure means that the polar ring signature appears \textcolor{black}{in the moment 1 map as a rotating disk surrounded by low-rotation gas.}  While the rotating disk can be clearly separated from the anomalous gas, the kinematic model is difficult to construct.  The low $S/N$ of the observations means that goodness-of-fit surface is quite flat, which makes distinguishing good fits from poor fits fairly difficult.  Nonetheless, we are able to generate a \textcolor{black}{model with} a similar polar ring signature in the moment 1 map as observed, although this model does have an abnormally low rotation velocity for the ring.  Like NGC 4632, we expect that a great improvement in the fitting can be made in future work by relaxing the circular and polarity assumptions.  In particular, slight deviations from polarity will allow the model ring to be more face-on, which will in turn lead to a higher model rotation velocity.  Given the starbursting nature of NGC 6156 as well as \textcolor{black}{its} status as a LIRG, it will be important to explore alternate explanations for the anomalous gas seen in WALLABY PDR1.  It may have been generated via (supernovae) outflows or some accretion event \textcolor{black}{that would produce a warp rather than a ring}.  \textcolor{black}{Given the importance of the galaxy inclination for the PRG interpretation coupled with the more face-on appearance of the stars, an investigation into the possibility of the gas being a warp is critical.  However, given the low $S/N$ and resolution of this data, untangling the full morphology of the gas will require both greater sensitivity and resolution.  Nonetheless, the regular structure of the anomalous gas as well as our ability to reproduce the salient features of the moment 1 map using our rudimentary model (namely the rotating disk signature surrounded by apparently face on gas) suggest that it is possible that NGC 6156 is a PRG, and that it should be considered a \textcolor{black}{potential \hi\ PRG} until deeper observations are available.}

%However, the regular structure seen in the WALLABY observations, as well as our ability to reproduce the salient features of the moment 1 map given our rudimentary model, suggests that the most likely explanation is that NGC 6156 is indeed a PRG.

Finally, we explored the detectability of \textcolor{black}{PRGs} in WALLABY by generating identical models with different observed geometries.  At low resolutions, it is nearly impossible to detect polar rings, while at high resolutions, \textcolor{black}{PRGs} are easily detected in either the moment 0 or moment 1 maps.  Combining this with the distribution of WALLABY PDR1 galaxy resolutions allows the estimation of the incidence rate of \textcolor{black}{PRGs}.  We find $\sim 1\% \le f_{i} \le 3 \%$, which is significantly higher than the canonical rate of $0.1\%$ as measured in \citet{Reshetnikov2011}.  \textcolor{black}{Our rate for multi-component \hi\ PRGs is in line with the incidence rate seen by \citet{Serra2012}, who found 3 PRGs around early type galaxies out of 166 observations.  It is worth noting here that we have not yet investigated whether there are any WALLABY detections where single component \hi\ detections are polar to stellar observations.  This rate is also similar to the rate of misaligned multi-component gas seen in \citet{Cao2022}.}

There are reasons to suspect that \textcolor{black}{PRGs} \textcolor{black}{in general, and \hi\ PRGs in particular,} are more common than previous stellar studies have suggested.  Firstly, the detection of stellar polar rings can be difficult due to their low surface brightness.  For instance, the faint overdensity detected around NGC 4632 in Figure \ref{Fig:NGC4632_StellarRing} is near the noise limit.  \hi\ gas has a much smaller dynamical range of surface brightnesses, making it easier to detect with new sensitive arrays.  Additionally, the interferometric nature of the WALLABY \hi\
provides the velocity structure which is where the key indications of NGC\,6156 being a \textcolor{black}{potential \hi\ PRG} are found.  Moreover, given that these two \textcolor{black}{potential \hi\ PRGs} are found \textcolor{black}{in \hi\, it is possible that} \hi\ rings have different formation mechanisms than stellar rings.  \textcolor{black}{It is worth noting here that most optical PRGs have been found around S0/E galaxies, while these structures are around spirals.  This might be due to a selection bias in detecting PRGs, or it may be a signature of a different origin.  Uncovering a larger sample of PRGs will help to determine whether there is real dependence on the host morphology for PRGs.}

One other potential reason for the increased incidence of \textcolor{black}{potential \hi\ PRGs} in WALLABY PDR1 is the environment.  The PDR1 fields were targeted to relatively nearby clusters and groups. Moreover, the NGC 4632 and NGC 6156 are at the outskirts of a massive  cluster and group respectively.  These environments have frequent tidal interactions and gas-rich mergers (relative to field galaxies), and both of these have the potential to generate \textcolor{black}{PRGs}.  Moving to the future, we will be able to investigate the incidence rate of \textcolor{black}{PRGs} in different environments as WALLABY will cover the majority of the southern sky.  This will remove a source of cosmic variance that may cause our large estimated incidence rate.

\textcolor{black}{Yet another possible reason for the higher incidence rate is that \hi\ rings may be an evolutionary stage in PRGs.  In \citet{Bekki1997}, simulations of PRG formation showed that star formation converted some of the gas in PRGs to stars.  It is possible that our \textcolor{black}{potential \hi\ PRGs} may eventually evolve into \textcolor{black}{stellar} PRGs (assuming that the morphology is correct).  However, simulations also show that polar structures are not permanent \citep{Khoperskov2021}.  If this evolution picture is correct, some PRGs would first form in \hi\, then convert to optical structures, and finally fall back to the galaxy disk.  In such a situation, it may be that the \hi\ rings last longer than optical rings, leading to a higher incidence rate.  To investigate this idea detailed observations of the optical counterparts to the anomalous gas will be necessary to determine the local star formation rates.}

If \textcolor{black}{PRGs} are indeed more common than previously estimated, WALLABY will find an abundance of new \textcolor{black}{PRGs}.  Predictions for the full WALLABY survey contained in \citet{Koribalski2020} indicate that $\ge10^4$ galaxies will be resolved by $\ge 5$ beams.  If the detection rate found in PDR1 holds true, this implies that WALLABY will detect hundreds of \textcolor{black}{PRGs}.  \textcolor{black}{Moreover, WALLABY plans to observe a subset of galaxies at a higher angular resolution of $\sim10\arcsec$.  This increased resolution means that most of those "postage stamps" will be resolved enough to see polar rings or other large scale kinematic misalignments.}  However, it is important to recognize that our \textcolor{black}{incidence rate estimate} is based on only two  \textcolor{black}{galaxies}, which is well within the realm of Poisson statistics.  As WALLABY moves into the future, this possibility will rapidly be resolved.  \textcolor{black}{Accurately measuring the rate of PRGs and the more general rate of extreme kinematic misalignments in gas is an interesting probe of galaxy formation.  Large scale cosmological simulations are producing these types of objects (e.g.  \citealt{Maccio2006,Roskar2010}), which will allow a comparison between the observed and simulated formation rates similar to the work of \citet{Elagali2018} on the formation of ring galaxies (where there is a stellar ring in the disk plane that are generally formed by drop through interactions). }

Regardless of \textcolor{black}{the precise PRG rate}, the two \textcolor{black}{potential \hi\ PRGs} found in WALLABY PDR1 are exciting and unique objects.  Unlike most \textcolor{black}{PRGs}, they have been detected in \hi\ rather than in stellar structures, they are gas rich, and we have been able to construct plausible kinematic models for the galaxies.   \textcolor{black}{As better observations and more sophisticated models are obtained for both galaxies, it will be possible to constrain the parameters of the ring progenitor (if they are indeed formed via mergers or flybys).  Understanding whether these are from interactions or gas accretion will provide constraints on galaxy formation and evolution.}

%\textcolor{black}{from TJ: this may be of interest, that we found, via vis inspection, two clear PRGs in a cosmo simulation of nearby uni from Neal Katz and Co., both of which were only seen with gas (versus stars);  i.e., they may be quite prevalent in high-res simulations;  you may want to ping Neal about this, although purely anecdotal at this time, it is provocative!).}

\section*{acknowledgement}

\textcolor{black}{We would like to thank the anonymous referee for their helpful comments and suggestions.}
Thanks to G. Meurer for useful comments. 

This scientific work uses data obtained from Inyarrimanha Ilgari Bundara / the Murchison Radio-astronomy Observatory. We acknowledge the Wajarri Yamaji People as the Traditional Owners and native title holders of the Observatory site. The Australian SKA Pathfinder is part of the Australia Telescope National Facility (\href{https://ror.org/05qajvd42}{https://ror.org/05qajvd42}) which is managed by CSIRO. Operation of ASKAP is funded by the Australian Government with support from the National Collaborative Research Infrastructure Strategy. ASKAP uses the resources of the Pawsey Supercomputing Centre. Establishment of ASKAP, the Murchison Radio-astronomy Observatory and the Pawsey Supercomputing Centre are initiatives of the Australian Government, with support from the Government of Western Australia and the Science and Industry Endowment Fund.

%The Australian SKA Pathfinder is part of the Australia Telescope National Facility (\href{https://ror.org/05qajvd42}{https://ror.org/05qajvd42}) which is managed by CSIRO. Operation of ASKAP is funded by the Australian Government with support from the National Collaborative Research Infrastructure Strategy. ASKAP uses the resources of the Pawsey Supercomputing Centre. Establishment of ASKAP, the Murchison Radio-astronomy Observatory and the Pawsey Supercomputing Centre are initiatives of the Australian Government, with support from the Government of Western Australia and the Science and Industry Endowment Fund. We acknowledge the Wajarri Yamatji as the traditional owners of the Observatory site.
    
WALLABY acknowledges technical support from the Australian SKA Regional Centre (AusSRC) and Astronomy Data And Computing Services (ADACS).

This research uses services or data provided by the Astro Data Lab at NSF's National Optical-Infrared Astronomy Research Laboratory. NOIRLab is operated by the Association of Universities for Research in Astronomy (AURA), Inc. under a cooperative agreement with the National Science Foundation.

This publication makes use of data products from the Wide-field Infrared Survey Explorer, which is a joint project of the University of California, Los Angeles, and the Jet Propulsion Laboratory/California Institute of Technology, funded by the National Aeronautics and Space Administration.

This work made use of the Inter-University Institute for Data Intensive Astronomy (IDIA) visualization lab\footnote{\href{https://vislab.idia.ac.za}{https://vislab.idia.ac.za}}. IDIA is a partnership of the University of Cape Town, the University of Pretoria and the University of the Western Cape. This work made use of the iDaVIE-v (immersive Data Visualisation Interactive Explorer for volumetric rendering) software (DOI - 10.5281/zenodo.4614116 - \href{https://idavie.readthedocs.io/}{https://idavie.readthedocs.io/} ) 

Parts of this research were supported by the Australian Research Council Centre of Excellence for All Sky Astrophysics in 3 Dimensions (ASTRO 3D), through project number CE170100013.

T.H.J. acknowledges support from the National Research Foundation (South Africa). AB acknowledges support from the Centre National d'Etudes Spatiales (CNES), France.  P.K. is supported by the BMBF project 05A20PC4 for D-MeerKAT.  K.S. acknowledges support from the Natural Sciences and Engineering Research Council of Canada (NSERC).

\section*{Data availability}

The WALLABY data are available through the \href{https://wallaby-survey.org/}{WALLABY data access portal} \textcolor{black}{at both the CSIRO ASKAP Science Data Archive (CASDA) and the Canadian Astronomy Data Center (CADC)}.  The galaxy models and modelling scripts will be shared upon request to the corresponding author.

%%%%%%%%%%%%%%%%%%%%%%%%%%%%%%%%%%%%%%%%%%%%%%%%%%

%%%%%%%%%%%%%%%%%%%% REFERENCES %%%%%%%%%%%%%%%%%%

% The best way to enter references is to use BibTeX:

%\bibliographystyle{mnras}
%\bibliography{example} % if your bibtex file is called example.bib

% Alternatively you could enter them by hand, like this:
% This method is tedious and prone to error if you have lots of references
\bibliographystyle{mnras}
\bibliography{PolarRingBib}

\appendix

\section{Warped and Ring Modelling}\label{Sec:KinModels}

\textcolor{black}{While this paper has focused on whether NGC 4632 and NGC 6156 are consistent with perfectly polar models, it is possible that they are extreme warps or inclined rings/disks.  In this appendix we explore this wider range of possible models.}

\subsection{`Warped' Models}\label{SubSec:WarpedModelling}

\textcolor{black}{As a part of the WALLABY PDR1 kinematic modelling efforts \citep{Deg2022}, both NGC 4632 and NGC 6156 were attempted to be modelled as `flat disks' (i.e. with constant geometries) using their proto-pipeline that combines both \barolo\ and \textsc{FAT}.  These models were deemed to be failures due to the kinematic misalignment between the inner disk and anomalous gas apparent in the data.  }

\textcolor{black}{Given the failure of the flat disk models, it is reasonable to use \barolo\ to fit a `warped' model for each galaxy.  In this approach, the inclination and position angle of the model are left as free parameters.  It is worth noting that these are sky plane coordinates and must be transformed to galaxy plane coordinates in order to determine the orientation of the anomalous gas with respect to the inner galaxy.  \barolo\ has a number of potential switches and modes that can be optimized for a particular modelling attempt.  In this case we used the `two-stage' mode, which fits the galaxy twice.  First it does a TR fitting of all the parameters of all rings separately. 
 Then it fits the resulting inclination and position angle profiles by a bezier function.  Those fitted geometric parameters are fixed in the second TR fitting of the data.  As such the final inclination and position angle parameters are not given with any uncertainties.   Nonetheless, the average uncertainties of the first fit can be used as approximate uncertainties on the inclination and position angle.}

\begin{figure*}
\centering
    \includegraphics[width=\textwidth]{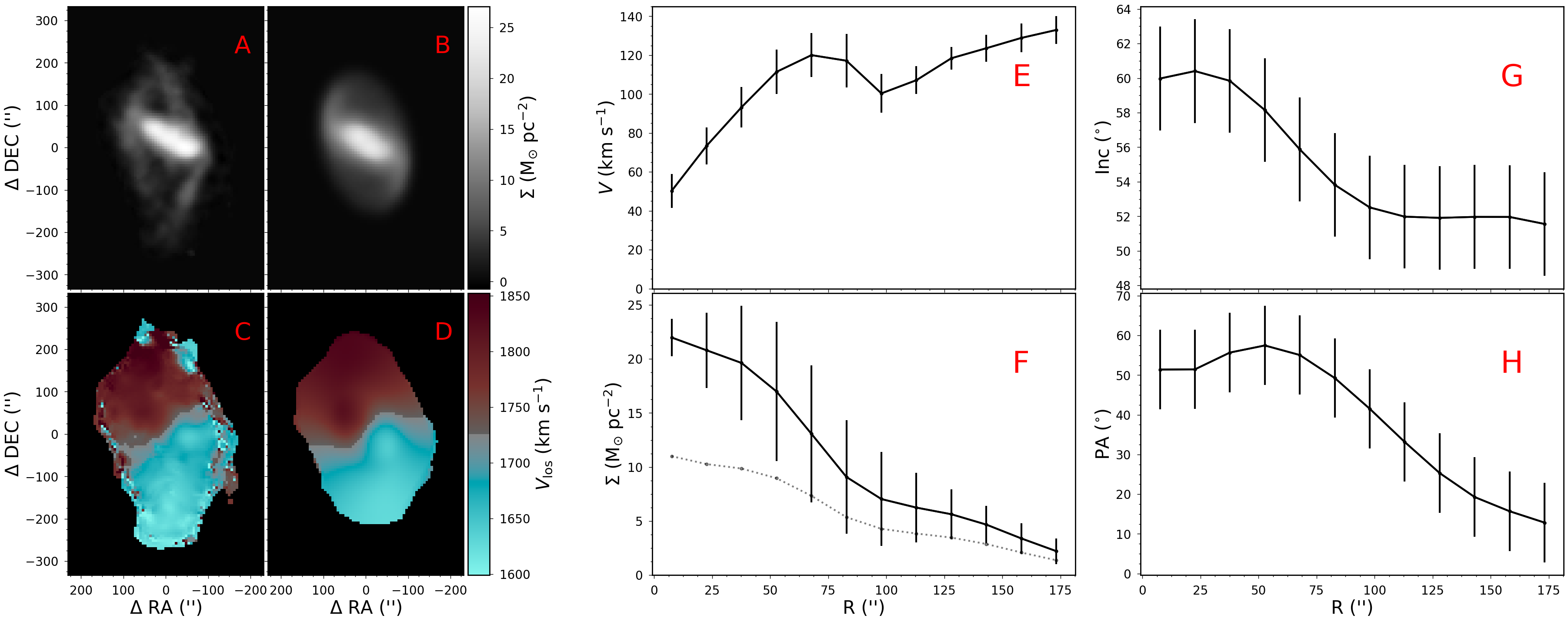} \caption{The best fitting \barolo\ warped model for NGC 4632.  Panels and lines are as in \ref{Fig:NGC4632Model} \textcolor{black}{with the exception of the dashed black line in Panel F.  This line shows the face-on surface density profile provided by the \barolo\ warped galaxy fit}.}
  \label{Fig:NGC4632_WarpedModel}
\end{figure*}

\textcolor{black}{Figure \ref{Fig:NGC4632_WarpedModel} shows the results of this two-stage warped \barolo\ fit to the data.  There are a number of key features to discuss in this fit.  Firstly, in this model, the inner orientation of the galaxy has $(i_{g},PA_{g})\approx(60^{\circ}, 55^{\circ})$, while the anomalous gas has an orientation of $(i_{r},PA_{r})\approx(52^{\circ},20^{\circ})$.  This orientation implies that that the anomalous gas is inclined at $30^{\circ}$ relative to the galaxy plane of NGC 4632. Such an orientation would make NGC 4632 a warped galaxy and not a PRG. However, this model does not reproduce the separation of the anomalous gas from the inner galaxy that is seen in panel A of the figure.  Additionally, the model velocity map shows an `S' shaped curve that roughly traces out the $V_{\rm{sys}}$ that is more pronounced than the observations.  Additionally, the systemic velocity appears to be slightly different from the galaxy's systemic velocity, as seen by the offset of the grey regions in the two moment 1 panels.}

\begin{figure*}
\centering
    \includegraphics[width=\textwidth]{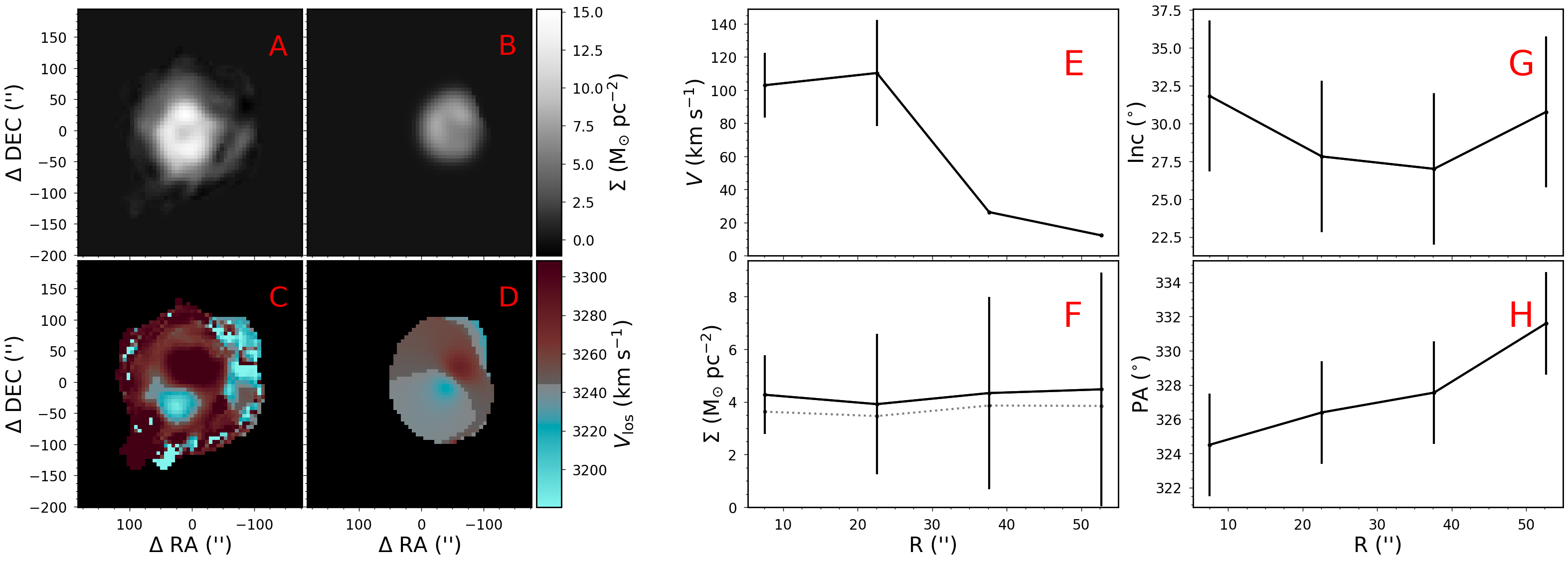} \caption{The best fitting \barolo\ warped model for NGC 6156.  Panels and lines are as in \ref{Fig:NGC4632_WarpedModel}. }
  \label{Fig:NGC6156_WarpedModel}
\end{figure*}

\textcolor{black}{A similar exercise can be done for NGC 6156.  Figure \ref{Fig:NGC6156_WarpedModel} shows the results of this analysis.  In this case \barolo\ has failed to model the anomalous gas.  The modelled systemic velocity appears to be slightly offset from the data as well. }

\subsection{Inclined Ring Models}\label{SubSec:InclinedModelling}

\textcolor{black}{The failure of the warped NGC 6156 model, as well as the fact that the warped NGC 4632 does not fully reproduce some key features of the anomalous gas, are chiefly due to the low $S/N$ and resolution of the data.  Complex models that consider the full parameter space of anomalous gas orientations are difficult to constrain with such data.  However, slightly less complex models can be used to answer more restricted questions.  One approach is to use \barolo\ to model the inner disk and anomalous gas separately as individual flat disk models.  This approach assumes that the anomalous gas can be described as a discrete ring or disk structure and does not consider warped models.}

\textcolor{black}{While such double flat disk models are simpler than the full \barolo\ models explored earlier, there are a number of important details to discuss.  Firstly, there is the separation of the inner galaxy from the anomalous gas.  For this work, we use the masks generated by the \textsc{iDaViE} separation discussed in Sec. \ref{Sec:WALLABY}.  These are seen in middle rows of Figs. \ref{Fig:NGC4632_WallabyOverview} and \ref{Fig:NGC6156_WallabyOverview}.  For both cases, a comparison of the middle (galaxy body) and bottom (anomalous gas) rows shows that the masking has isolated the main bodies of the galaxies quite well.  In particular, for NGC 6156, the apparently Gaussian profile has been transformed into a more standard, albeit noisy, flat/double-peaked profile.}

\textcolor{black}{Upon separation of the galaxy into the inner and outer components, a first `flat-disk' fit to the inner disk is required. This fit starts with an estimate of the initial parameters, but, when the data has a low $S/N$ and resolution like both our galaxies, the initial estimates on the parameters have a strong influence on the final model.  In particular the initial estimate for the inclination provides strongly affects the results.  For this particular analysis, we have adopted an optical inclination of $\sim62.5^{\circ}$ based on the HSC image.  For this inner disk fit, all geometric parameters as well as the rotation curve and surface density profile are fit by \barolo\ using the \textit{two-stage} fitting option.  This particular fitting option fits the galaxy twice, and in the second fit, the geometric parameters are fixed based on the results of the first fitting step.  In order to keep the separate fits for the inner and outer disk consistent, the outer disk center point and systemic velocity are fixed to the results of the inner disk `flat-disk' model.}

\begin{figure*}
\centering
    \includegraphics[width=\textwidth]{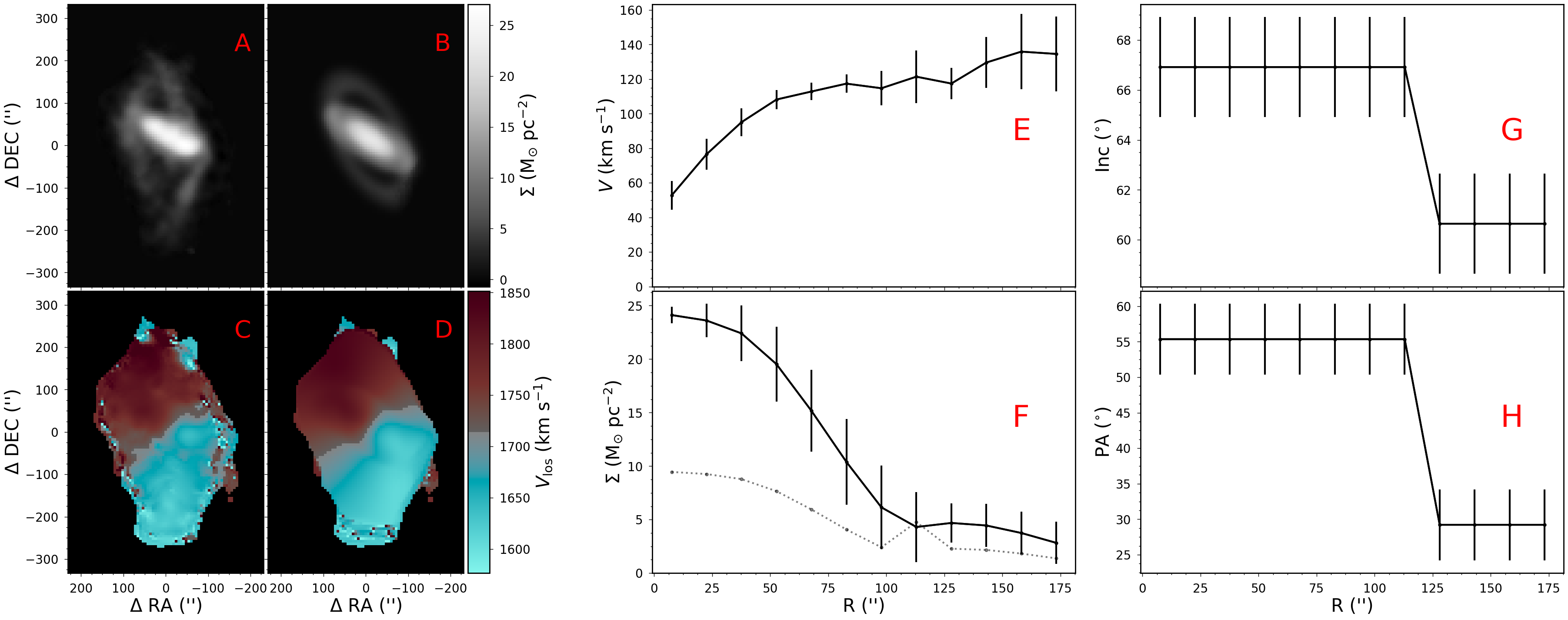} \caption{A combined model from individual flat disk \barolo\ fits to the inner disk and anomalous components of NGC 4632. Panels and lines are as in \ref{Fig:NGC4632_WarpedModel}.}
  \label{Fig:NGC4632_InnerOuter}
\end{figure*}

\textcolor{black}{Figure \ref{Fig:NGC4632_InnerOuter} shows the results for a combined model from individual flat disk fits to the inner galaxy and anomalous gas components using \barolo.  In this case, the fit to the anomalous gas is clearly distinct from the inner disk, which the warped was unsuccessful at replicating. In this model, the outer structure is inclined at $\sim24^{\circ}$ with respect to the inner disk.  However, the moment 0 map of the model shows that the shape and extent of the outer model does not match the observed structure of the anomalous gas. And, while the moment 1 map no longer shows the `S' shape for the systemic velocity, a separation between the different components velocity structures can be seen that is not as clear in the actual observations.}

\textcolor{black}{Given the results of these combined fit for NGC 4632 as well as the poorer $S/N$ and resolution of the NGC 6156 data, we have chosen not to generate a combined model for NGC 6156.  One of the potential reasons for these poor results is a particular quirk of \barolo.  When \barolo\ is passed a mask, whether it be the WALLABY \textsc{SoFiA} masks used for the warped models or the inner and outer masks for the combined models, \barolo\ still runs its source finding algorithm.  Rather than examining the full cube, it restricts the source finding to the masked region.  This works well for the initial problems that \barolo\ was designed to address, but it causes issues here.  It is likely that the \barolo\ source finding is not extracting the full content of the anomalous gas (although it is likely working quite well on the inner disk), leading to poor fits for both the warped and combined models.}

\textcolor{black}{Comparing the results of these more general modelling approaches for both NGC 4632 and NGC 6156 to the purely polar models of Sec. \ref{Sec:PRGConsistency}, it is clear that there all three types of models have the ability to reproduce the data.  Given the issues in each fit, we are unable to definitively determine the morphology of the anomalous gas.  Thus, even though our observations are \textit{consistent} with PRG models, we cannot rule out warped or inclined ring models.  Without improved observations, we must continue to classify these as \textit{potential} PRGs.}

\section{\KS{Perfectly Polar Ring Geometry}}\label{Sec:Appendix_PRMath}

\KS{This appendix contains a derivation of the perfectly polar ring relations in Equations~\ref{Eq:IncREq} and Eq. \ref{Eq:ThetaREq} that underpin the kinematic models in Sec.~\ref{SubSec:PRGFormalism}. They arise from} the transformation of points between three coordinate systems; the sky plane, the galaxy plane, and the ring plane.  These transformations are accomplished using rotations where the rotation angles are related to physical quantities.

In the most general sense, the transformation of points in galaxy plane to the sky plane is
\begin{equation}
    \mathbf{x}_{s}=R_{z}(\theta_{g})R_{x}(i_{g})\mathbf{x}_{g}~,
    \label{Eq:GalaxyToSky}
\end{equation}
where $\mathbf{x}_{s}$ are the sky-plane coordinates, $\mathbf{x}_{g}$ are the galaxy plane coordinates, $R_{x}(i_{g})$ is a rotation about the $x_{g}$ axis by the inclination $i_{g}$, $\theta_{g}=PA_{g}+90^{\circ}$, and $R_{z}(\theta_{g})$ is a rotation about the $z_{g}$ axis by $\theta_{g}$.  
A similar transformation will convert points in the ring plane to points in the sky plane 
\begin{equation}
    \mathbf{x}_{\textcolor{black}{s}}=R_{z}(\theta_{r})R_{x}(i_{r})\mathbf{x}_{r}~,
\end{equation}
\KS{where the model polar ring is assumed to have the same geometric center as the galaxy body.}  Filling in the matrices, the transformation becomes
\begin{equation}
    \mathbf{x}_{s}=
    \left(
    \begin{matrix}
    c_{\theta r} & -s_{\theta r}c_{ir} & s_{\theta r}s_{ir}\\
    s_{\theta r} & c_{\theta r}c_{ir} & -c_{\theta r}s_{ir}\\
    0 & s_{ir} & c_{ir}
    \end{matrix}
    \right)
    \mathbf{x}_{r}~,
    \label{Eq:RingRotMatrix}
\end{equation}
where $c$ and $s$ are the $\cos$ and $\sin$ functions respectively, and the subscript indicates the variable \textcolor{black}{(i.e. $c_{\theta r}=\cos(\theta_{r})$)}.

Alternatively, it is possible to first transform points in the ring plane to the galaxy plane and then to the sky plane.  A point in a polar ring can be transformed to galaxy plane coordinates via
\begin{equation}
    \mathbf{x}_{g}=R_{z}(\beta)R_{y}(\phi)R_{x}(90^{\circ})\mathbf{x}_{r}~.
    \label{Eq:GalaxyToRing}
\end{equation}
In this equation, we have introduced an extra rotation of $\phi$ about the $y$ axis.  This rotation is required for the further rotation to the sky plane coordinates.  However, it is confusing from the point of view \KS{of a rotating circular ring of uniform density, since that coordinate is morphologically degenerate}.  If the polar ring were to have an overdensity at some location, then $\phi$ would control where that overdensity appears on the sky.  

\KS{Combining Equations~\ref{Eq:GalaxyToSky}~and~\ref{Eq:GalaxyToRing} gives}
\begin{equation}
    \mathbf{x}_{s}=R_{z}(\theta_{g})R_{x}(i_{g}) R_{z}(\beta)R_{y}(\phi)R_{x}(90^{\circ})\mathbf{x}_{r}~.
\end{equation}
This can be simplified slightly by setting $\theta_{g}=0$ to give
\begin{equation}
    \mathbf{x}_{s}=R_{x}(i_{g}) R_{z}(\beta)R_{y}(\phi)R_{x}(90^{\circ})\mathbf{x}_{r}~.
\end{equation}
This simplification does not lose any generality as the galaxy's position angle can simply be added to the ring's position angle once $\theta_{r}$ is calculated for $\theta_{g}=0^{\circ}$.  Filling in the rotation matrices gives
\begin{equation}
    \mathbf{x}_{s}=
    \left(
    \begin{matrix}
    c_{\beta}c_{\phi} & c_{\beta}s_{\phi} & s_{\beta}\\
   c_{ig}s_{\beta}c_{\phi}+s_{ig}s_{\phi} & c_{ig}s_{\beta}s_{\phi} -s_{ig}c_{\phi} & -c_{ig}c_{\beta}\\
    s_{ig}s_{\beta}c_{\phi}-c_{ig}s_{\phi} &s_{ig}s_{\beta}s_{\phi} +c_{ig}c_{\phi} & -s_{ig}c_{\beta}
    \end{matrix}
    \right)
    \mathbf{x}_{r}~.
    \label{Eq:GalaxyBetaRotMatrix}
\end{equation}
Equating Eq. \ref{Eq:RingRotMatrix} to Eq. \ref{Eq:GalaxyBetaRotMatrix} gives a set of relations that can be used to solve for $\theta_{r}$, $i_{r}$, and $\phi$.  This implies that $\phi$ is not a free rotation.  It, like $\theta_{r}$ and $i_{r}$ is uniquely determined by the $(i_{g},\theta_{g},\beta)$ angles.  Using the [3,3] elements of the matrices (third row, third column), we get
\begin{equation}
    c_{ir}=-s_{ig}c_{\beta}~.
\end{equation}
Similarly, dividing the [1,3] elements by the [2,3] elements gives
\begin{equation}
    t_{\theta r}=\frac{s_{\beta}}{c_{\beta}c_{ig}}~.
\end{equation}
In this equation, we have not simplified the $\beta$ terms in order to allow $\theta_{r}$ to vary across $0^{\circ}-360^{\circ}$. 

A key takeaway for this derivation is that a polar ring only has a specific range of allowed inclinations and position angles for a given galaxy body.  For instance, for a face-on galaxy, the ring must be edge-on but can be at any position angle (determined by $\beta$).  Conversely, when a galaxy is edge-on, the polar ring must have a position angle that lies at $90^{\circ}$ to the galaxy's position angle, but it can be face-on or edge-on depending on the $\beta$ angle.

% Don't change these lines
\bsp	% typesetting comment
\label{lastpage}
\end{document}